\documentclass{optica-article}
\journal{opticajournal} % for journals or Optica Open
\articletype{Research Article}

\usepackage[utf8]{inputenc}

\usepackage{hyperref}
\usepackage{amsmath}
\usepackage{amsfonts}
\usepackage{graphicx}
\usepackage{physics}
\usepackage{siunitx}
\usepackage{verbatim}
\usepackage{listings}
\usepackage{subfigure}

\usepackage{color}
\usepackage{comment}
\usepackage{todonotes}
\usepackage{soul}
\newcommand{\nl}{\nonumber \\}
\newcommand{\naturals}{\mathbb{N}}

\newcommand{\operator}[1]{\ensuremath{\hat{#1}}}
\renewcommand{\op}[1]{\operator{#1}}

\renewcommand{\vector}[1]{\vectorbold{#1}}
\renewcommand{\v}[1]{\ensuremath{\vector{#1}}}
\newcommand{\uv}[1]{\ensuremath{\underbar{\v{#1}}}}

\newcommand{\fn}[1]{\ensuremath{\scalebox{0.75}{(#1)}}}

\newcommand{\ft}{\fn{t}}

\newcommand{\ftr}{\fn{t,\position}}
\newcommand{\ftrv}{\fn{t,\v\position}}
\newcommand{\fcart}{\fn{x,y,z}}

\newcommand{\polarizationAxis}{\uv{e}}

\newcommand{\conjugate}[1]{#1^{*}}
\renewcommand{\i}{\ensuremath{\mathbf{i}}}
\newcommand{\intensity}[1]{\ensuremath{\abs{#1}^{2}}}

\newcommand{\e}{\ensuremath{e}}

\newcommand{\dt}{\dv{t}}
\newcommand{\pdt}{\pdv{t}}

\newcommand{\integralInfinite}{\ensuremath{\int_{-\infty}^{+\infty}}}
\newcommand{\intInf}{\integralInfinite}

\newcommand{\hermite}[2]{\ensuremath{\text{H}_{#1}\fn{#2}}}

\newcommand{\position}{\ensuremath{r}}
\newcommand{\transverse}[1]{\ensuremath{#1_{\text{T}}}}
\newcommand{\positionTransverse}{\ensuremath{\transverse{\v\position}}}

\renewcommand{\time}{\ensuremath{t}}
\renewcommand{\t}{\time}
\renewcommand{\t}{\time}

\newcommand{\velocity}{\ensuremath{v}}

\newcommand{\force}{\ensuremath{F}}
\newcommand{\hc}{\text{H.c.}}
\newcommand{\annihilation}{\ensuremath{\operator{a}}}
\newcommand{\ann}{\annihilation}
\newcommand{\creation}{\ensuremath{\operator{a}^{\dagger}}}
\newcommand{\cre}{\creation}

\newcommand{\rotating}[1]{\ensuremath{\tilde{#1}}}
\newcommand{\rot}[1]{\rotating{#1}}

\newcommand{\energyOp}{\operator{H}}
\newcommand{\Hamiltonian}{\energyOp}
\newcommand{\Ham}{\Hamiltonian}

\newcommand{\pauli}{\operator{\sigma}}

\newcommand{\pauliZ}{\pauli_{z}}
\newcommand{\pauliP}{\pauli^{\dagger}}
\newcommand{\pauliM}{\pauli}

\newcommand{\ef}{\ensuremath{E}}

\newcommand{\permittivity}{\ensuremath{\epsilon}}
\newcommand{\permittivityOfFreeSpace}{\permittivity_{0}}
\newcommand{\pfs}{\permittivityOfFreeSpace}

\newcommand{\wavelength}{\ensuremath{\lambda}}
\newcommand{\wl}{\wavelength}
\newcommand{\wavenumber}{\ensuremath{k}}
\newcommand{\wn}{\wavenumber}

\newcommand{\angularFrequency}{\ensuremath{\omega}}
\newcommand{\af}{\angularFrequency}

\newcommand{\profile}{\ensuremath{f}}

\newcommand{\modeindex}{\ensuremath{n}}
\newcommand{\coupling}{\ensuremath{g}}
\newcommand{\phase}{\ensuremath{\phi}}

\newcommand{\width}{\text{w}}
\newcommand{\waist}{\ensuremath{\width_0}}

\newcommand{\HG}[1]{\ensuremath{\text{HG}_{#1}}}

\newcommand{\dipolemomentElectric}{\ensuremath{d}}
\newcommand{\dmE}{\dipolemomentElectric}

\newcommand{\rabi}{\Omega}

\newcommand{\decay}{\ensuremath{\gamma}}
\newcommand{\mode}[1]{#1_{\modeindex}\fn{\wavenumber}}
\newcommand{\modes}[2]{#1_{\modeindex_{#2}}\fn{\wavenumber #2}}
\newcommand{\modeTime}[1]{#1_{\modeindex}\fn{\time,\wavenumber}}
\newcommand{\modeTimes}[2]{#1_{\modeindex_{#2}}\fn{0,\wavenumber #2}}

\newcommand{\modeProfile}{\profile_{\modeindex}\fn{\transverse{\v\position}}}
\newcommand{\modeTransverse}[1]{#1_{\modeindex}}
\newcommand{\crosssection}{\ensuremath{\mathcal{A}}}

\newcommand{\rotDrive}[1]{\rot{\rot{#1}}}
\newcommand{\annRot}{\rot{a}}
\newcommand{\creRot}{\rot{a}^{\dagger}}
\newcommand{\creRotDrive}{\rotDrive{a}^{\dagger}}

\newcommand{\pauliZRot}{\rot{\sigma}_{z}}
\newcommand{\pauliPRot}{\rot{\sigma}^{\dagger}}
\newcommand{\pauliMRot}{\rot{\sigma}}

\newcommand{\pauliMRotDrive}{\rotDrive{\sigma}}

\newcommand{\afThresh}{\ensuremath{\angularFrequency_{th,\modeindex}}}

\newcommand{\atomic}[1]{#1_{\text{A}}}
\newcommand{\afDiff}{\Delta_{\af\wn, A}}
\newcommand{\afDiffs}[1]{\Delta_{\af\wn #1, A}}
\newcommand{\afDiffAtomic}{\Delta_{\af A,\wn }}
\newcommand{\afDiffAtomicDrive}{\Delta_{\af A,\text{dr} }}

\newcommand{\lossFiber}{\kappa}

\newcommand{\amplitude}{\alpha}
\newcommand{\drive}[1]{\ensuremath{#1_{\text{dr}}}}
\newcommand{\rabiTransverse}{\rabi\fn{\positionTransverse}}
\newcommand{\rabiDrive}{\drive{\rabi}\fn{\positionTransverse}}

\renewcommand{\phase}[1]{\ensuremath{\Phi_{\modeindex}\fn{#1}}}
\newcommand{\Phase}{\ensuremath{\Phi_{\modeindex}\fn{\v\position, \wn}}}
\newcommand{\Phases}[1]{\ensuremath{\Phi_{\modeindex #1}\fn{\v\position, \wn #1}}}

\newcommand{\drivePhase}{\ensuremath{\Phi_{\text{dr}}\fn{t, \v\position}}}

\newcommand{\drivePhaseP}{\ensuremath{\Phi_{\text{dr}}\fn{\v\position}}}

\newcommand{\tzero}{\ensuremath{t_{0}}}

\newcommand{\noise}{\ensuremath{\op{\xi}_{\kappa}}}
\newcommand{\noiseRot}{\ensuremath{\rot{\xi}_{\Gamma}}}
\newcommand{\noiseRotDrive}{\ensuremath{\rotDrive{\xi_{\Gamma}}}}
\newcommand{\statewidth}{\ensuremath{\Gamma_{\text{F}}}}
\newcommand{\statewidthMode}{\ensuremath{\Gamma_{\text{F}\modeindex}}}
\newcommand{\statewidthThZero}{\ensuremath{\Gamma_{\text{F,0}}}}
\newcommand{\statewidthTh}{\ensuremath{\Gamma_{\text{F,th}}}}
\newcommand{\lightshift}{\ensuremath{\Delta_{\text{F}}}}
\newcommand{\stateLine}{\zeta}
\newcommand{\stateLineP}{\stateLine\fn{\positionTransverse}}

\newcommand{\driveForce}{\ensuremath{\rotDrive{F}}}

\newcommand{\vAtom}{\atomic{\v\velocity}}
\newcommand{\polarization}{\modeTransverse{\polarizationAxis}\fn{\positionTransverse}}
\begin{document}

\title{Atoms in hollow-core fibers: A QED approach}

\author{
Thomas W. Clark,\authormark{1,*}
Luca Vincetti,\authormark{2}
Peter Domokos,\authormark{1}
}

\address{\authormark{1}Wigner Research Centre for Physics, Budapest H-1121, Hungary\\
	\authormark{2}Department of Engineering “Enzo Ferrari,” University of Modena and Reggio Emilia, I-41125 Modena, Italy
}
\email{\authormark{*}thomas.clark@wigner.hun-ren.hu}
\date{}
% \maketitle

\begin{abstract}
	We outline mechanical effects of light-matter interaction inside hollow-core optical fibers. Starting with quantized electromagnetic radiation, we demonstrate how dispersion, mode functions and losses define an open quantum system and how subsequent Langevin equations can be used to predict spatially-dependent vacuum forces. Conceptually, we reveal new, geometry-induced, forces that have no equivalence in unbounded 3-D space and, practically, show how the general spatial dependence can be greatly approximated by free-space Ince-Gaussian modes: such that the forces can be described analytically. By also considering the effects of drive and fluctuations, we provide an extensive overview of both control and cooling within the limitations of a 2-level atomic system.
\end{abstract}

\section{Introduction}
With the introduction of photonic crystal fibers (PCF)\cite{knight_all-silica_1996, cregan_single-mode_1999}, specifically those that exploit photonic band-gaps (PBG)\cite{birks_endlessly_1997, deutsch_photonic_1995}, optical confinement could be considered outside the limits of total internal reflection, and outside the limits of traditional geometries. In fact, without internal reflection, the central glass medium is not needed at all \cite{knight_photonic_1998,temelkuran_wavelength-scalable_2002}. With this new-found void however, it wasn't long before the concept of atomic gas-filled cells was introduced\cite{benabid_compact_2005} and thus, intrafiber light-matter interaction became a reality: with laser cooling and ultracold atomic trapping following afterwards\cite{bajcsy_laser-cooled_2011,christensen_trapping_2008}. Therefore, although optical forces have historically been observed since Kepler\cite{kepler_cometis_1619} and  systematized since Maxwell and Ashkin\cite{maxwell_treatise_1873, ashkin_acceleration_1970}, we outline how this new type of confinement changes things: such that, within the limits of low saturation and a two-level atom, there is now the potential for miniaturising cold atomic laboratories.

This progress from cold atom laboratory to room-temperature cell did not happen in a vacuum, so to speak.  When considering optical atomic forces more generally, preeminent in the literature are the first stopping of an atomic beam \cite{phillips_laser_1982}, three-dimensional cooling \cite{chu_three-dimensional_1985} and three-dimensional trapping \cite{raab_trapping_1987}.  Practical milestones, like shortening the distance required for cooling  \cite{soding_short-distance_1997} were also necessary, as was, of particular relevance to this work,  the introduction of geometrical constraints: primarily, efficient cavity cooling \cite{alge_all_1997, horak_cavity-induced_1997, domokos_semiclassical_2001, maunz_cavity_2004, murr_large_2006}. Additionally,  the current paper was preceded by work that allowed ultra low light atom-optical interaction within a fiber \cite{ghosh_low-light-level_2006}, atomic torque outside fibers \cite{le_kien_light-induced_2006} and intrafiber spectroscopy \cite{yang_atomic_2007}. The trapping of ultracold atoms  within and the efficient guiding of cold atoms throughout a hollow core system were particular milestones \cite{christensen_trapping_2008, vorrath_efficient_2010}, as was the loading of laser-cooled atoms \cite{bajcsy_laser-cooled_2011, langbecker_highly_2018}. With the subsequent in-fiber cooling of atoms and molecules \cite{sommer_laser_2019, wang_enhancing_2022} and even the beginnings of waveguide QED with `giant' atoms \cite{terradas-brianso_ultrastrong_2022}, the modern relevance of models for fiber light-atom interaction is clear.

In this context,  we consider the full QED (quantum electrodynamical) approach, where both light and atoms are confined quantum mechanically. This follows two early theoretical works that highlighted the direct impact of geometry on known cooling effects. The first, highlights a mechanism for cavity-enhanced friction and Doppler cooling, even in directions perpendicular to a resonator's axis \cite{domokos_anomalous_2004}. The second, postulates a new type of cooling whose origins can be directly attributed to a fiber-like geometry: dubbed a geometric resonance \cite{szirmai_geometric_2007}. The current paper, explores this idea of geometric resonance, but where we must now consider the modal decomposition in some depth. The radiative decomposition, while powerful for describing forces on atoms in real space, is much more involved in bounded geometries. In fact, a complete set of modes is, typically, unknown in analytical form, except for some specialised, idealised, boundary conditions. Instead, we will demonstrate how an elliptical decomposition of the radiation field, even without knowledge of the individual spatial functions, leads to an efficient description of atomic dipole radiation inside a hollow-core fiber. Efficient, in that we arrive at analytically tractable forms of the various optical forces, using only some generic characteristics of the modes.

\section{A quantum electrodynamic model}
\subsection{Fiber-confined radiation}
Fibers, by nature, are strongly geometric entities, confining the electromagnetic field in two dimensions, while leaving it free to propagate along the fiber axis, $z$. From the outset therefore, we expect a coordinate separation similar to that for free space, i.e. a transverse eigenvalue equation for modes with wavenumber $\wn_{z}$ and angular frequency $\af$:
\begin{equation}
	\label{eq:transverseHelmholtz}
	\nabla_T^2 f(\positionTransverse) = - \Big(n^2(\positionTransverse) \frac{\af^2}{c^2}- \wn_{z}^2\Big) f(\positionTransverse).
\end{equation}
This assumes that there is no divergence of the field and that the index of refraction, $n(\positionTransverse)$, and the speed of light, $c$, are defined conventionally.

Whereas free space solutions are still relevant to the fiber core (\autoref{sec:mode-functions}), it is the boundary with the cladding that defines a fiber. Such is the case, there are a variety of physical techniques to inhibit this coupling, all of which carry different refractive profiles, $n(\positionTransverse)$, and thus solutions to the eigenfuctions, $f(\positionTransverse)$, and their respective eigenvalue spectra, $\wn_T^2 \equiv \af^2/c^2-\wn_z^2$. In general, the spectrum of the transverse momentum, $\wn_T$, is continuous, but organized into discrete branches according to the number of nodal lines in the core, $n$. For weak confinement, i.e., when a large part of the mode function, $f(\positionTransverse)$, expands into the dielectric part of the fiber, the eigenvalues also depend on $\af$. In this case, there is no simple analytical expression for the \emph{dispersion relation} between the optical angular frequency $\af$ and the propagation constant, $\wn_{z}$. The stronger the confinement however, the weaker the dependence of the eigenvalue on $\af$. Ultimately, in the perfect confinement limit, where the penetration of the radiation field into the cladding material is negligible, the eigenvalues of Eq.~(\ref{eq:transverseHelmholtz}) form a discrete set and are determined solely by the geometry of the core.  In such a case, the discrete set of eigenvalues represent \emph{threshold} frequencies, $\afThresh$, which give the minimum energy for each allowed branch of propagating modes at $\wn_{z}=0$. The simplified mode dispersion relation can than be expressed by
\begin{equation}
	\label{eq:dispersion_relation}
	\mode\af = \sqrt{\afThresh^{2} + c^{2}\wn^{2}},
\end{equation}
where, for convenience, we now abbreviate the axial wavenumber, $\wn_{z}$, by $\wn$ throughout. This is usually considered a good model for metallic waveguides with little optical penetration (\emph{skin} effect). 

This can also be taken as an appropriate analytical approximation for PCF hollow core fibers in the strongly confining limit even if the genuine dispersion relation may be different in principle, as the eigenvalues of Eq.~(\ref{eq:transverseHelmholtz}) slightly depend on the optical frequency $\af$. For propagating modes far above the threshold, $\wn\gg \afThresh/c$, the usual linear dispersion relation is retained. Note that this is the typical case at the optical frequencies close to atomic resonances. In the critical region, for modes close to threshold $\wn\rightarrow 0$, the spectrum of modes and the actual form of the dispersion relation cannot be resolved because these modes are spectrally broadened. The fiber loss, $K$, is defined as the extinction rate per unit length and the decay of fiber mode amplitudes per unit time is thus influenced by propagation, 
\begin{equation}
	\label{eq:loss}
	\mode\lossFiber = \frac{K \omega}{k}.
\end{equation}
That is, the damping rate of modes near the threshold ($k\rightarrow 0$, $\af \approx \afThresh$) diverges.  

With this property of linewidth broadening of long-wavelength modes, the approach based on modes strongly (perfectly) confined into the hollow core can be used consistently in the full range of propagation wavenumbers. The exponential decay of the field away from the core within the dielectric cladding, on the scale of the wavelength, can be considered a loss of perfectly confined modes, and can be taken into account via a single extinction rate $K$. A threshold angular frequency can be defined for each branch of modes on the basis of geometry. Far above threshold, the decay rate of the modes approaches a constant: $\kappa \rightarrow cK$, whereas the long-wavelength modes are very lossy, $\kappa \approx K \afThresh / k \rightarrow \infty$. Much like the definition of modes within a linear Fabry-P\'erot resonator, the scaling of the loss rate defines then the extent to which discrete eigenvalues become Lorentzian spectra, which are ultimately smeared together in the limit of weak confinement. Analogously, the continuous spectrum of eigenvalues of Eq.~(\ref{eq:transverseHelmholtz}) near a threshold is conceived as a broadened discrete resonance. 

Fiber losses are generally low, \emph{i.e.} on the order of $\unit{dB/km}$, and are effectively negligible for short fibers. For simplicity, we also assume $K$ to be independent of the transverse branch index, $\modeindex$. This assumption is in line with the analogy of hollow core confinement with resonances in a linear cavity. In this latter case, the linewidth is close to constant for longitudinal modes in a range where the reflectivity of the mirrors does not depend significantly on the wavelength.

\begin{figure}[tb]
	\centering
	\includegraphics[width=\linewidth]{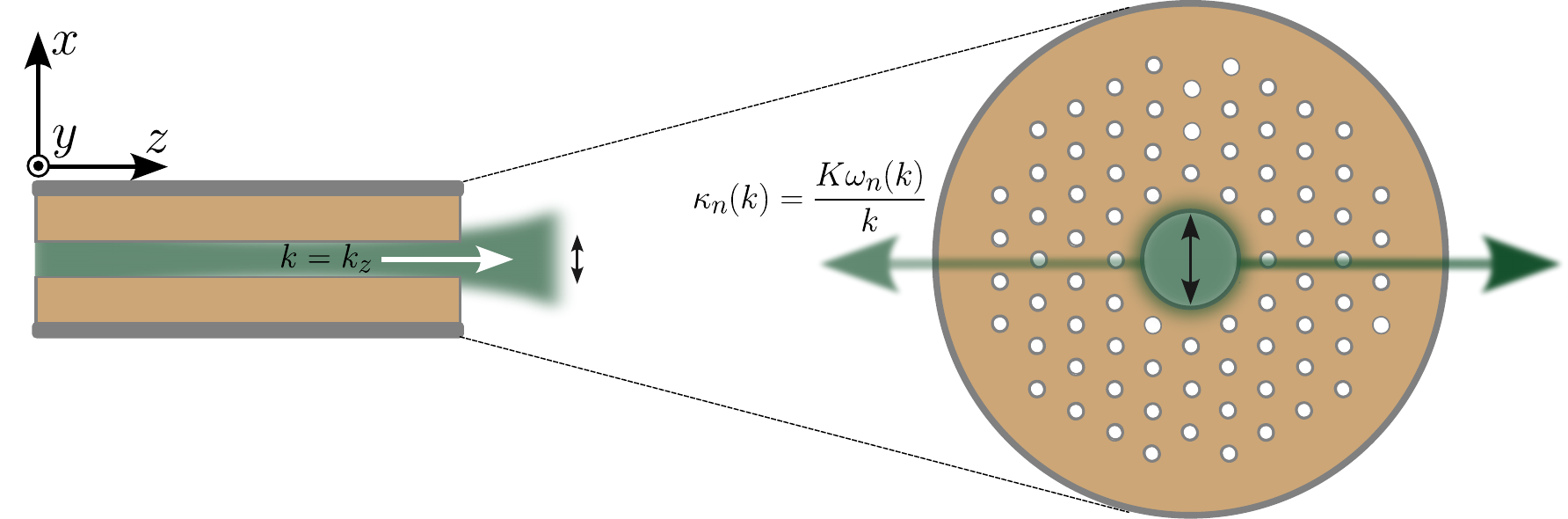}
	\caption{ \label{fig:fiber-geometry}
		An overview of a photonic crystal fiber geometry and the associated loss mechanism.
	}
\end{figure}

Loss plays an important role in the theory, as a natural regularization on both physical and mathematical grounds (\autoref{fig:fiber-geometry}). This is most relevant close to a threshold, where $\omega_n(k)\rightarrow \omega_{n, \text{th}}$, and so the wavenumber vanishes, $k\rightarrow 0$. These long wavelength modes then have diverging phase velocity and loss rate but vanishing group velocity following from \autoref{eq:dispersion_relation}. Thus, a narrow bandwidth light pulse, scattered off of an atom into these modes, does not propagate but sticks to the atom, creating an unusual domain of interaction: one worthy of investigation.

Once the light is separated into confined (transverse) and propagating (longitudinal) components, where modes are naturally discrete in two dimensions and continuous in $z$, annihilation and creation operators can be associated with the modes:
\begin{equation}
	\label{eq:dispersion}
	[\ann_{n}(\wn_{z}), \cre_{n'}(\wn_{z}')] = \delta(\wn_{z}-\wn_{z}')\delta_{n,n'}.
\end{equation}
Here, $n$ indexes each transverse mode, and the mode annihilation and creation operators, $\ann$ and $\cre$ are functions of $\wn_z$ and, unusually, carry units of  $1/\sqrt{[\wn]} = \sqrt{m}$.
The electric field modes follow naturally:
\begin{align}
	\op{\v\ef}\ftr & = \op{\v\ef}^{+}\ftr + \op{\v\ef}^{-}\ftr                                                                                                                                    \\
	               & = \i\sqrt{\frac{\hbar}{4\pi\pfs}}  \sum_{n} \modeProfile\polarization \intInf \mode{\ann} \sqrt{\frac{\mode\af}{\modeTransverse\crosssection}} e^{ \i\Phase}~\dd{\wn} + \hc,
\end{align}
where $\profile$, $\positionTransverse$, $\uv e$ and $\crosssection$ represent the transverse mode profile, transverse coordinate, polarisation unit vector and cross-sectional area, respectively. Furthermore, we define the mode cross section by
\begin{equation}
	\modeTransverse\crosssection = \int\dd{\positionTransverse}^{2} \abs{\modeProfile}^{2},
\end{equation}
where $\modeProfile$ or $\crosssection$ are peak normalized or normalized to `1', respectively, depending on the mode functions used. We also choose to carry the complex components in the exponential phase, where $\Phase = \phase{\positionTransverse, z} + \wn z$ (real) for a general plane wave. A summary of the geometric details are presented in \autoref{fig:fiber-geometry}.

%\subsubsection{Mode functions}
\subsection{Mode functions}
\label{sec:mode-functions}
\begin{figure}[b]
	\centering
	\includegraphics[width=\linewidth]{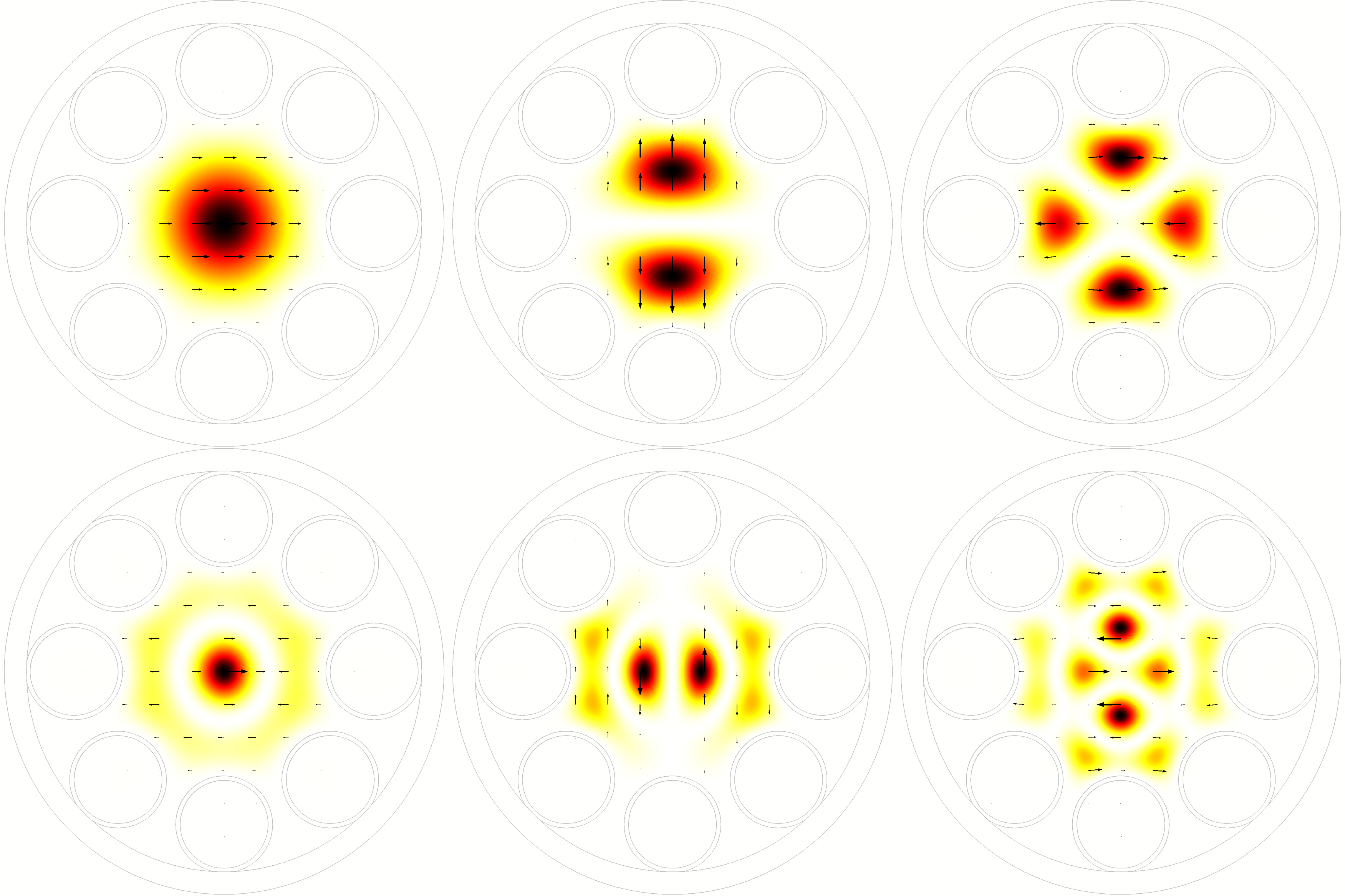}
	\caption{A selection of realistic, numerically computed, fiber mode profiles. From top left to bottom right: $LP_{0,1}$, $LP_{1,1}$, $LP_{2,1}$, $LP_{0,2}$, $LP_{1,2}$, $LP_{2,2}$.}
	\label{fig:profiles--sim}
\end{figure}
The outline above is still somewhat abstract. What exactly are the mode profiles of hollow-core fibers? Unfortunately, in a general hollow-core system, differing manufacturing processes, surface roughness and other local details make a general answer prohibitively complicated. Using finite element analysis however (COMSOL with anisotropic Perfectly Matched Layer)\cite{cucinotta_perfectly_1999}, we can gain insight into the most relevant mode profiles.

Computationally, we assume the electric field to carry a standard mode profile:
\begin{equation}
    \v{E}(\v{r}_T,z)=\modeProfile\polarization e^{-(K_n+\i \wn_{z_n})z}
\end{equation}
where $K_n$ and $k_{z_n}$ are the mode loss and wavenumber, respectively, with common units (1/m). The mode properties are obtained by numerically solving the so called curl-curl equation:
\begin{equation}
    \v{\nabla}\times(\v\nabla\times\v{E})-n^2(\v{r}_T)\left(\dfrac{\omega}{c}\right)^2\v{E}=0,
\end{equation}
as obtained by decoupling the Maxwell equations. By applying a variational finite element procedure, the following algebraic eigenvalue equation is obtained \cite{selleri_complex_2001}:
\begin{equation}
    \left( 
    [A] - \left( (K+ik_z)\dfrac{c}{\omega} \right)^2 [B]
    \right)\left\{ F \right\}=0.
    \label{eq:eigen_eq}
\end{equation}
Here, the eingenvector, $\{F\}$, is the discretized field vector, that is the numerical approximation of $\modeProfile\polarization$, and $A$ and $B$ are sparse and symmetric matrices. Solving this equation, we obtain the mode profiles, a selection of which are presented in \autoref{fig:profiles--sim}.

\begin{figure}[tb]
	\centering
	\includegraphics[width=\linewidth]{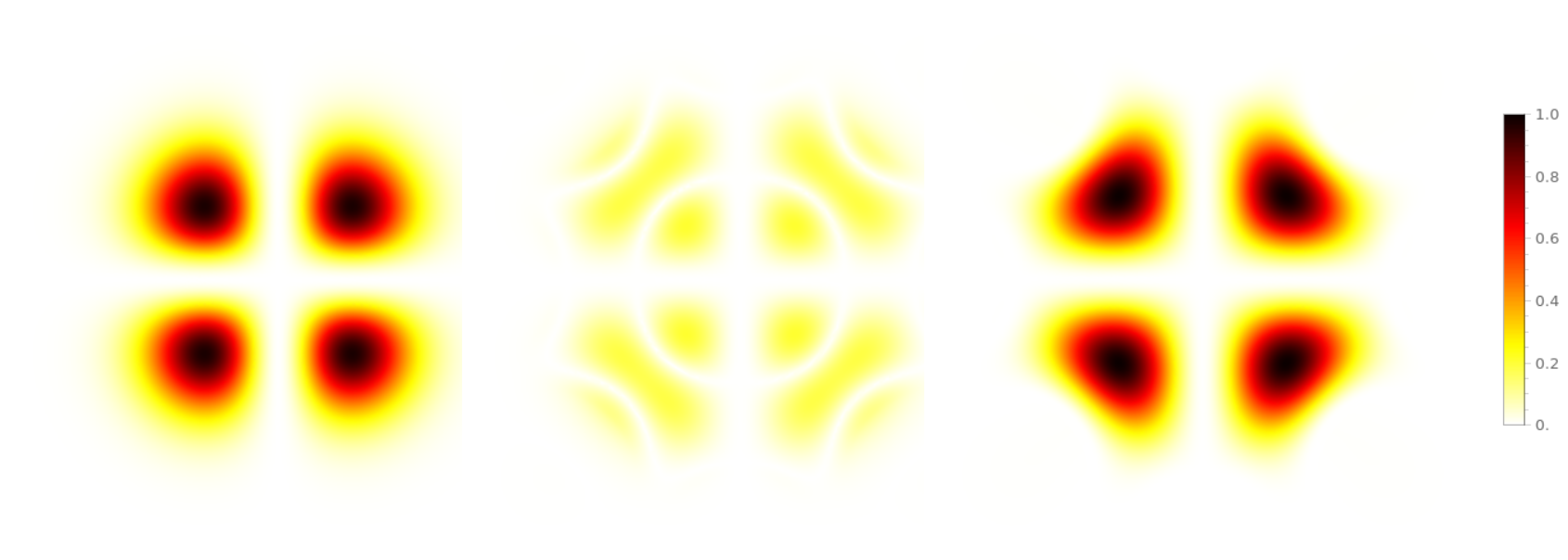}
	\caption{A free-space modal approximation, absolute residual and simulated intensity profile on the same colour scale.}
	\label{fig:diff--simulated-analytical}
\end{figure}

\begin{figure}[tb]
	\centering
	\includegraphics[width=\linewidth]{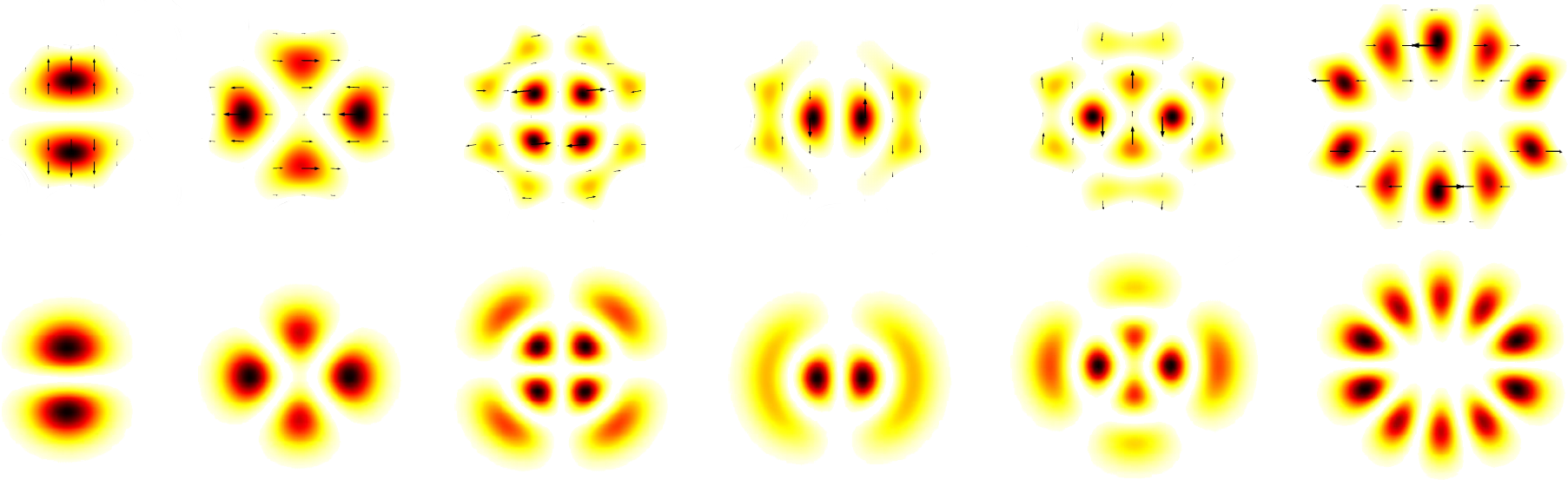}
	\caption{A comparison of the numerically predicted (upper) and analytically modelled (lower) mode profiles. From left to right, the IG mode numbers and ellipticity are (1,0;1.4), (2,2;0.35), (4,2;0.1), (3,2;0.4), (4,2;0.35) and (5,5; 1.5). }
	\label{fig:profiles--comparisons}
\end{figure}

Here, we soon see strong deviation from the patterns of traditional, step-index fibers, which are often modelled with Bessel beams. In contrast, the simulated profiles resemble beams more commonly seen while propagating freely, albeit without expansion. In fact, the first few modes can be very well approximated using Hermite-Gaussian (HG) functions (\autoref{fig:diff--simulated-analytical}). Other modes however are more obviously elliptical, rather than rectangular or circular, and so, to treat the fiber mode profiles concisely, we consider the Ince-Gaussian (IG) family \cite{bandres_incegaussian_2004} (\autoref{fig:profiles--comparisons}).

\begin{figure}[bt]
	\centering
	\includegraphics[width=\linewidth]{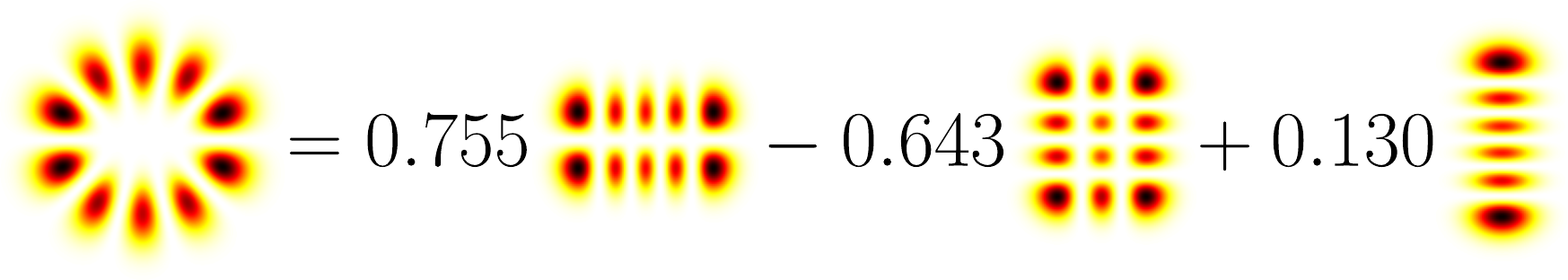}
	\caption{A seemingly complex, elliptically symmetric, mode can be concisely expressed by a superposition of only three rectangularly symmetric (HG) modes.}
	\label{fig:superposition}
\end{figure}

This family however, is not well known and can be more difficult to work with: which can lead to lengthy programming. On the other hand, their modes can be broken down into finite, linear superpositions of the more familiar Hermite-Gaussian ones, bounded by the mode number (physically, this constraint comes from the need for a Gouy phase), $N = l + m$:
\begin{equation}
	\text{IG}^{\sigma}_{N} = [T^{\sigma}_{N}]\text{HG}^{\sigma}_{N},
\end{equation}
where $\sigma \in {o[\text{dd}],e[\text{ven}]}$ is the parity and $T$ is a numerical transformation matrix. Such is the case, individual IG beams can be  expanded as a limited sum of HG beams:
\begin{equation}
	\label{eq:IG-expansion}
	\text{IG}^{\sigma}_{N,m} = \sum_{l,m} T_{m,l}\HG{l,m}^{\sigma}: \forall l \in \naturals_{0}, \forall m \in \naturals_{\sigma}, N=l+m,
\end{equation}
\textit{e.g.}
\[IG^{\text{o}}_{5,5} = 0.755 \HG{4,1} -0.643 \HG{2,3}+ 0.130 \HG{0,5} ~(\text{\autoref{fig:superposition}}).\]

We can therefore reasonably assume that the mode profile has the form
\begin{equation}
	\modeTransverse\profile\fn{\positionTransverse} = \sum_{l,m}^{l+m} \alpha_{n,l,m} \HG{l,m}\fn{\positionTransverse},
\end{equation}
such that \(\alpha\) is an arbitrary constant, either complex most generally or real in the main. Such a model significantly simplifies subsequent calculations, as the modes' spatial derivatives can now be extracted with simple recurrence relations, \textit{i.e.}
\begin{equation}
	\label{eq:gradient-field-profile}
	\grad\modeTransverse\profile\fn{\positionTransverse} =
	\sum_{l,m} \alpha_{n,l,m} \begin{pmatrix}
		\frac{2}{\waist}(\sqrt{l} \HG{l-1,m} - \frac{x}{\waist}\HG{l,m}) \\
		\frac{2}{\waist}(\sqrt{m} \HG{l,m-1} - \frac{y}{\waist}\HG{l,m})
	\end{pmatrix}.
\end{equation}
Unfortunately, there is a cost to this approach, as the cross-section of these modes varies with increasing mode number and so this has to be taken into account \autoref{fig:waist--mode-number}. For fuller derivations, expansions and comments on the scaling, see the relevant supplementary material (\autoref{sec:supplementary}).

\subsection{Fiber-confined atoms}

Fiber confinement does not simply affect the bare radiation: in our system, there are atomic gases and so spontaneous emission too. The Hamiltonian, under electric dipole and rotating wave approximations, can then be written:
\begin{align}
	\label{eq:hamiltonians}
	\Ham                 & = \Ham_{\text{field}} + \atomic\Ham + \Ham_{\text{dipole}},                                                                                         \\
	\Ham_{\text{field}}  & = \hbar \sum_{n} \intInf\dd{\wn} \mode\af \modeTime\cre \modeTime\ann,                                                                              \\
	\atomic\Ham          & = \hbar \atomic\af \pauliZ,                                                                                                                         \\
	\Ham_{\text{dipole}} & = -\i\hbar \left( \pauliM+\pauliP \right) \sum_{n} \intInf \modeTime\ann \left(  \modeProfile \mode\coupling   e^{ \i\Phase} \right) \dd{\wn} + \hc \\
	                     & \simeq \i\hbar \sum_{n} \intInf \modeTime\cre \pauliM \left(\modeProfile \mode\coupling   e^{-\i\Phase} \right)\dd{\wn} + \hc,
\end{align}
where the approximation leading to the last line corresponds to the rotating-wave approximation, \(\atomic\af\) is the atomic frequency, we have used the two-level Pauli operators and where $g$ is the electric dipole coupling. This is expressed by
\begin{equation}
	\mode\coupling  = \frac{\dmE}{\sqrt{4\pi\pfs\hbar}} \sqrt{\frac{\mode\af}{\modeTransverse\crosssection}},
\end{equation}
where \(\dmE\) is the electric dipole moment along the field polarization and we have assumed aligned dipole and field axes, in accordance with a linear polarizability model of the atom. The coupling has the dimension \([\mode\coupling] = \frac{\sqrt{m}}{s}\). Furthermore, the dipole moment can be, more intuitively, expanded in terms of the natural linewidth, \(\decay\):
\begin{equation}
	\mode\coupling^{2} = \frac{\decay c}{4\pi} \frac{\atomic\crosssection}{\modeTransverse\crosssection}\frac{\mode\af}{\atomic\af}
\end{equation}
by using the resonant atomic cross section, $\atomic\crosssection =\frac{3\wavelength_{A}^{2}}{2\pi}$.

\subsection{Heisenberg--Langevin equations of motion}
With the basic model above,  we use Heisenberg--Langevin-like equations to model a driven, open fiber system with mode-dependent loss:
%From the Heisenberg equation of motion,
%
% \begin{equation}
%   \dt \op{o} = \frac{1}{\i \hbar}[\op{o}, \Ham],
% \end{equation}
%
%\begin{subequations}
% \begin{align}
%   \dt \mode\ann = [\mode\ann, \Ham]/\i\hbar  = &-\i\mode\af\modeTime\ann +  \pauliM \Big( \modeProfile \mode\coupling \Big)  e^{-\i \Phase} \text{ and }\\
%   \dt \pauliM = [\pauliM, \Ham]/ \i\hbar = &-\i\atomic\af\pauliM + 2\pauliZ \sum_{\modeindex}\intInf\dd{\wn}  \mode{\ann} \Big( \modeProfile \mode\coupling \Big) \e^{+\i \Phase} .
% \end{align}
%\end{subequations}
%
%Therefore, with losses,
\begin{subequations}
	\label{eq:HeisenbergLangevin}
	\begin{align}
		\dt \mode{\ann} \ft & = -(\i\mode\af + \mode\lossFiber)\mode\ann + \pauliM\ft \Big( \modeProfile \mode\coupling \Big) e^{-\i \Phase} + \noise, \\
		\dt \pauliM\ft      & = -\i\atomic\af \pauliM\ft
		+ 2\pauliZ \sum_{\modeindex}\intInf\dd{\wn} \mode{\ann} \Big( \modeProfile \mode\coupling \Big) \e^{+\i \Phase} \,,
	\end{align}
\end{subequations}
where the quantum noise terms, $\noise$, represent the fluctuations accompanying the fiber loss with rate $\mode\lossFiber$.
The first equation can be formally integrated,
\begin{align}
	\label{eq:field-evolution}
	\ann_{\modeindex}\fn{\wn,\t} =
	%& \e^{-(\i\mode\af + \mode\lossFiber) \t}\left( \ann\fn{\wn,\tzero,\v\position} + \int_{\t0}^{\t} \e^{(\i\mode\af + \mode\lossFiber) \tdash} \pauliM\fn{\tdash} \modeProfile \mode\coupling e^{-\i\Phase } \dd{\tdash} \right) \nl
	\modeTransverse\ann\fn{\wn,\tzero}\e^{-(\i\mode\af + \mode\lossFiber) (\t-\tzero)}
	 & + \modeProfile\mode\coupling \, e^{-\i\Phase}
	\int_{0}^{\t - \tzero}
	\pauliM\fn{\t - $\tau$} \e^{-(\i\mode\af + \mode\lossFiber) \tau}
	\dd{\tau} \nl
	 & +  \underbrace{\int_{0}^{\t - \tzero} \dd{\tau} \noise (\t-\tau) \e^{-(\i\mode\af + \mode\lossFiber) \tau}}_\text{noise},
\end{align}
and then simplified, as the last term is integrated noise that plays no role in subsequent calculations (assuming normal order products at zero temperature). Furthermore, without loss of generality, the starting time can be zeroed: $\tzero=0$ and the formal solution inserted into Eq.~(\ref{eq:HeisenbergLangevin}b), such that the atomic evolution is described by an integro-differential equation that can be treated within the Markovian (forgetful) approximation. Anticipating this approximation, we first move into the interaction frame, \( \pauliM = \pauliMRot \e^{-\i\atomic\af \t} \):
\begin{align}
	\dt \pauliMRot\ft & \simeq
	2\pauliZRot \sum_{\modeindex}  \intInf\dd{\wn} \Big( \modeProfile\mode\coupling \Big)^{2}
	\int_{0}^{\t}
	\pauliMRot\fn{\t - $\tau$} \e^{ - \left[\i(\mode\af -\atomic\af) + \mode\lossFiber\right] \tau}
	\dd{\tau} \nl
	                  & + \underbrace{2\pauliZRot \sum_{\modeindex}\intInf\dd{\wn} \modeProfile\mode\coupling e^{+\i \Phase -\left[\i (\mode\af - \atomic\af) + \mode\lossFiber\right] \t} \modeTransverse\annRot\fn{\wn,\tzero}}_\text{noise}.
\end{align}
Here, the first term has a mean value of zero and so, in practical terms, can also be conceptualized as noise. The last term contains much of the essential physics and can be simplified further. Since the spectral structure of the fiber modes form a one-dimensional broadband continuum around the atomic resonance frequency, the integral over $k$, in the exponential $\e^{\i(\atomic\af - \mode\af) \tau}$, vanishes outside a very narrow time window around $\tau=0$. In the interaction picture therefore, the atomic polarization does not change during this short period, so $\pauliMRot\fn{\t - $\tau$} \approx \pauliMRot\fn{\t}$ can be taken out of the temporal integral, and we can use $2\pauliZRot\fn{\t}\cdot \pauliMRot\fn{\t} \equiv - \pauliMRot\fn{\t}$.
The temporal integral can then be carried out, leaving two integrals over $\wn$. Finally, we also apply the low-saturation limit in accordance with the linear polarizability model of the atom, ($\pauliZ \approx -1/2$) and so, in the spirit of the Markovian approximation, the evolution of the atomic polarization can be approximated by a single quantum Langevin equation with a broadening and shift in resonance due to the fiber modes:
\begin{subequations}
	\begin{equation}
		\label{eq:evolution--vacuum}
		\dt\pauliMRot \simeq -\pauliMRot \left[  \statewidth\fn{\positionTransverse}  + \i \lightshift\fn{\positionTransverse} \right]
		-\noiseRot\fn{\t, \v\position} \text{ and }
	\end{equation}
	\begin{equation}
		\label{eq:statewidth}
		\statewidth\fn{\positionTransverse} =  \sum_{\modeindex}\intInf\dd{\wn} \frac{\modeProfile^{2}\mode\coupling^{2}}{\mode\lossFiber^{2} + \afDiffAtomic^{2}}\mode\lossFiber \,,
	\end{equation}
	\begin{equation}
		\label{eq:lineshift}
		\lightshift\fn{\positionTransverse} = \sum_{\modeindex}\intInf\dd{\wn} \frac{\modeProfile^{2}\mode\coupling^{2}}{\mode\lossFiber^{2} + \afDiffAtomic^{2}} \afDiffAtomic \,,
	\end{equation}
	\begin{equation}
		\label{eq:noise}
		\noiseRot\ftrv = \sum_{\modeindex}\intInf\dd{\wn} \mode\annRot\fn{\wn,\tzero} \Big( \modeProfile\mode\coupling \Big) e^{+\i \Phase + (\i\afDiffAtomic - \mode\lossFiber) \t}\,.
	\end{equation}
\end{subequations}
The noise has a mean of zero, and its second-order correlation function can be approximated by a unit pulse (i.e., white noise):
\begin{equation}
	\label{eq:NoiseCorrelation}
	\langle \noiseRot(t_1, r) \noiseRot^\dagger(t_2,r) \rangle \approx 2\statewidth\fn{\positionTransverse}  \, \delta(t_1-t_2)\;,
\end{equation}
where the coefficient of the unit pulse can be directly derived from the definition of the fluctuation operators Eq.~(\autoref{eq:statewidth}) by using Eq.~(\autoref{eq:dispersion}), and the short correlation time (unit pulse) reflects the Markovian approximation. The simplicity of the final result is in accordance with the dissipation-fluctuation theorem, which is encoded in the close relation of the forms of the dissipation and noise in Eqs.~(\autoref{eq:statewidth}) and (\autoref{eq:noise}), respectively.

%\begin{equation}
% D_\Gamma  = \int_0^{\infty} dt 2 {\text Re}\left\{ \langle \noiseRot\ftrv \noiseRot(0,r)^\dagger \rangle \right\} = 
%\sum_{\modeindex}\intInf\dd{\wn} \frac{\modeProfile^{2}\mode\coupling^{2}}{\mode\lossFiber^{2} + \afDiffAtomic^{2}}\mode\lossFiber = 2\statewidth\fn{\positionTransverse} 
%\end{equation}
%where we used Eq.~(\ref{eq:dispersion}) to eliminate one of the integrals and summations.

%
\begin{figure}
	\centering
	\subfigure{\includegraphics[width=0.32\textwidth]{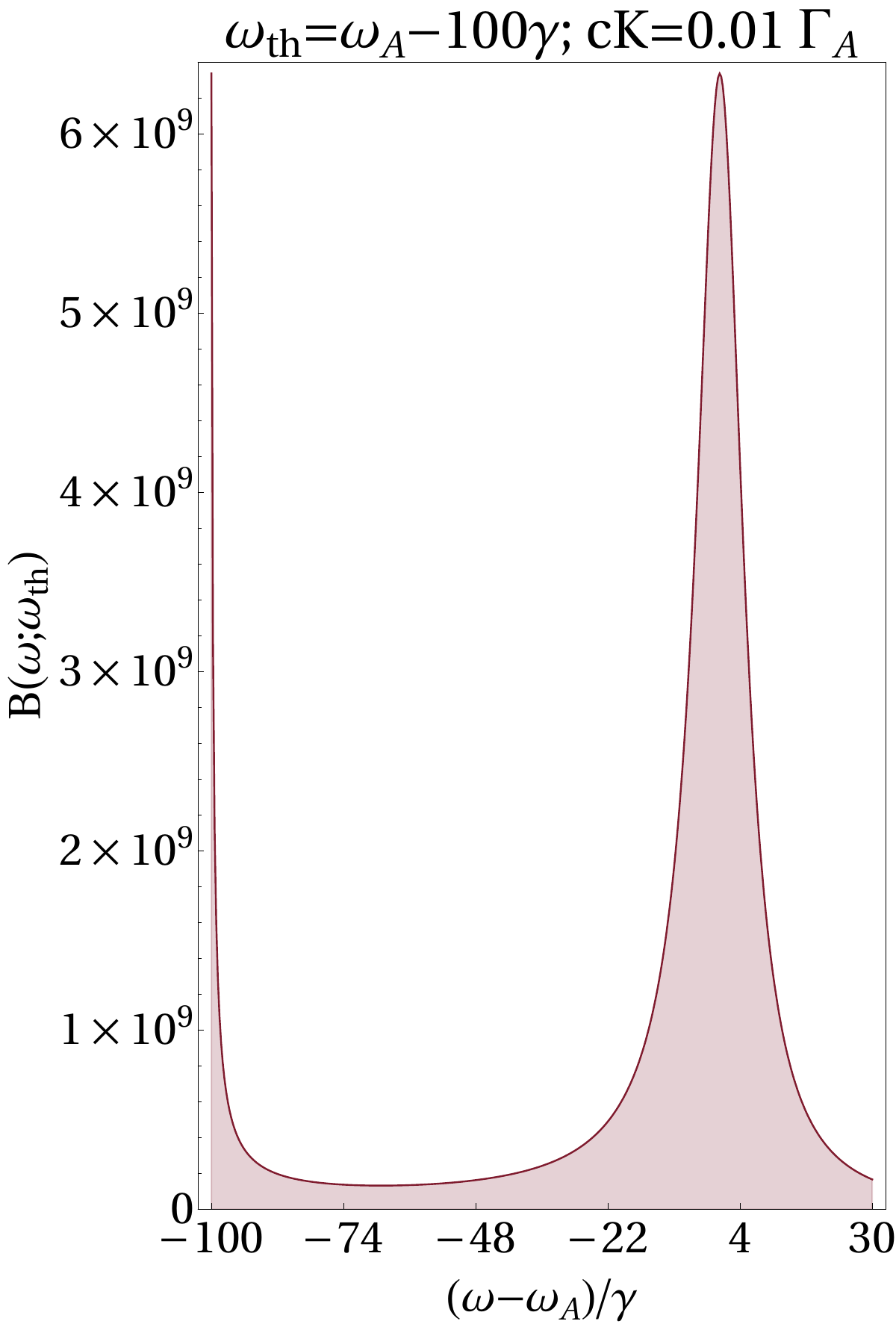}}
	\subfigure{\includegraphics[width=0.32\textwidth]{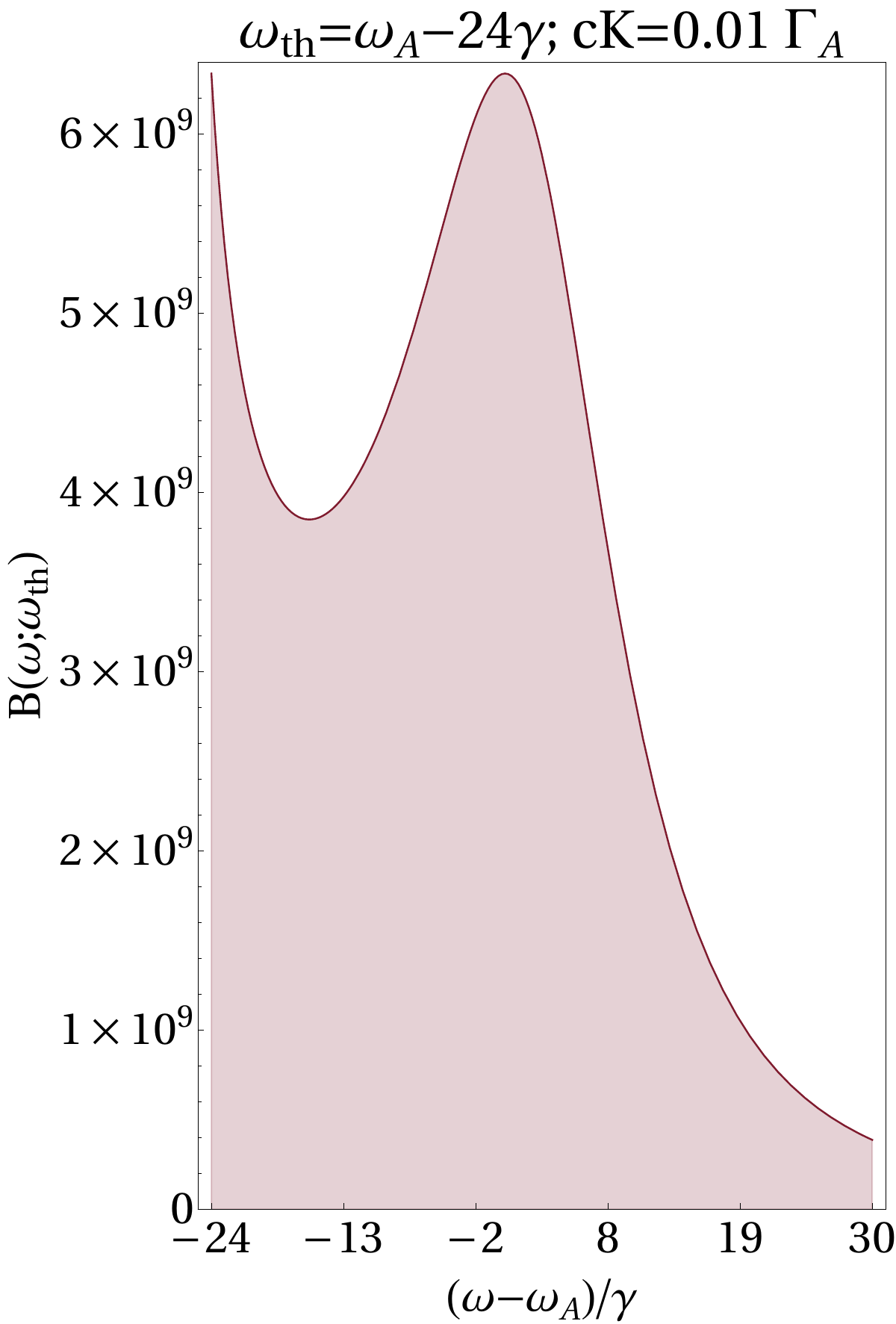}}
	\subfigure{\includegraphics[width=0.32\textwidth]{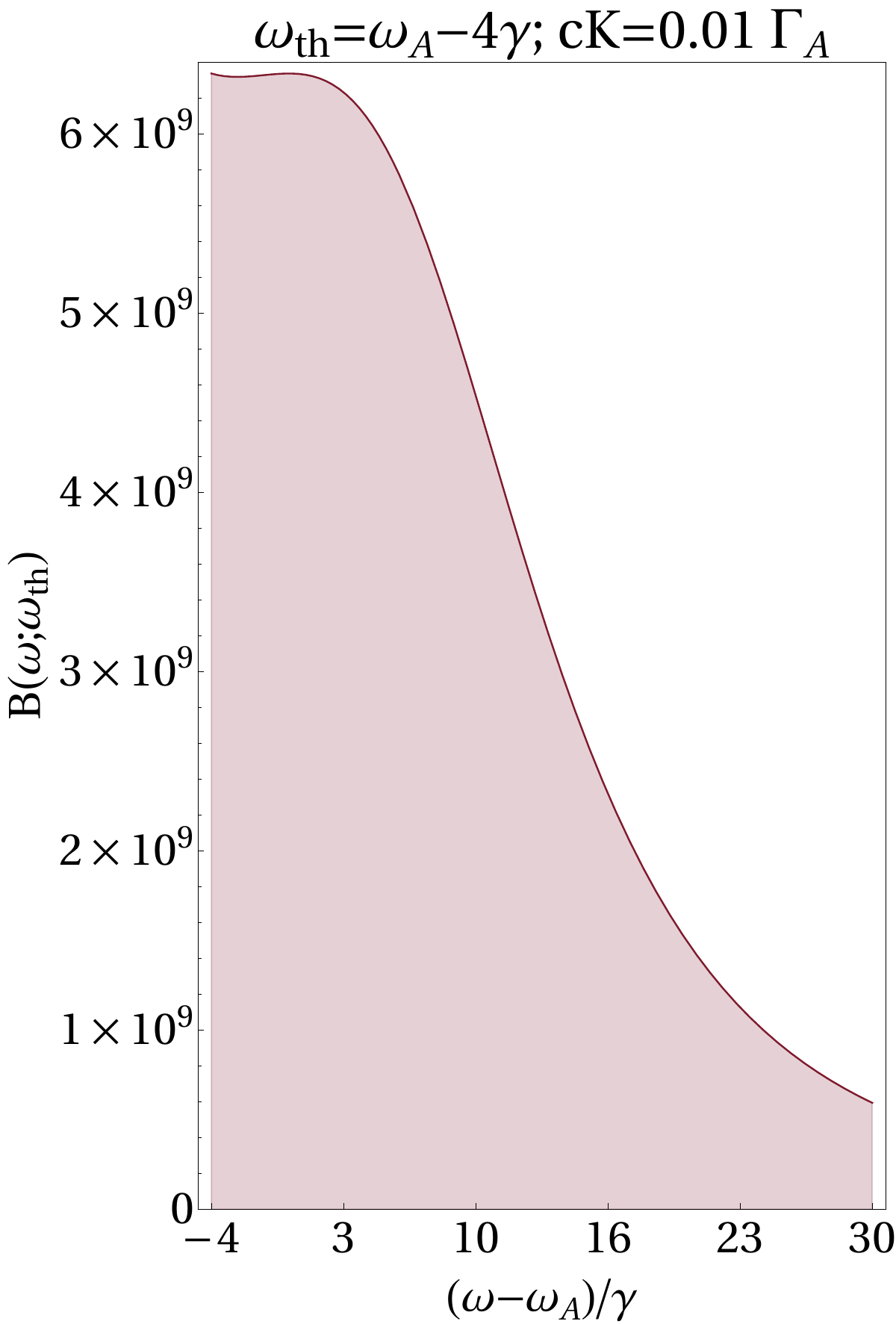}}
	\caption{
		An overview of the integrand, $B$, of the vacuum-induced linewidth broadening (\autoref{eq:broadening--integrand}) as the mode threshold frequency approaches an atomic resonance. Here, the function is plotted within the future limits of integration, and the area under the curve highlighted, for illustration. The value of $cK$ is chosen for illustration and $\atomic\af=1$ for numerical convenience.
	}
	\label{fig:broadening--integrand}
\end{figure}
\section{Fiber QED}
\label{sec:fiber-physics}
Although \autoref{eq:evolution--vacuum} may seem familiar to those who work with QED models -- and the expectation is that for large enough area of the hollow core, the rate of spontaneous emission is close to that in free space -- we subscript them here because they carry unique properties in the context of confined geometries. Considering our definition of fiber dispersion, \autoref{eq:dispersion},
\begin{align}
	\statewidth\fn{\positionTransverse} & =  2\sum_{\modeindex}\modeProfile^{2} \int_{0}^{\infty}\dd{\wn} \mode\coupling^{2} \frac{\mode\lossFiber}{\mode\lossFiber^{2} + \afDiffAtomic^{2}}\nl
	                                    & = 2\sum_{\modeindex}\modeProfile^{2} \int_{0}^{\infty}\dd{\wn} \mode\coupling^{2} \frac{K\mode\af/k}{(K\mode\af/\wn)^{2} + (\atomic\af - \mode\af)^{2}} \nl
	                                    & = 2\frac{\decay}{4\pi}\frac{\atomic\crosssection}{\atomic\af}   \sum_{\modeindex} \frac{\modeProfile^{2}}{\crosssection_{\modeindex}} \int_{0}^{\infty}\dd{\wn}   \frac{c K \wn (\afThresh^{2} + c^{2}\wn^{2})}{K^{2}(\afThresh^{2} + c^{2}\wn^{2}) + \wn^{2}(\atomic\af - \sqrt{\afThresh^{2} + c^{2}\wn^{2}})^{2}}.
\end{align}
Or, in terms of angular frequencies:
\begin{align}
	\statewidth\fn{\positionTransverse}
	 & = \frac{\decay}{2\pi}\frac{\atomic\crosssection}{\atomic\af}   \sum_{\modeindex} \frac{\modeProfile^{2}}{\crosssection_{\modeindex}}
	\int_{\afThresh}^{\infty}\dd{\af_{n}} [\frac{\af_{\modeindex}}{c \sqrt{\af_{\modeindex}^{2} - \afThresh^{2}}}]
	\frac{c K \frac{1}{c}\sqrt{\af_{n}^{2}-\afThresh^{2}} \af_{n}^{2}}{K^{2} \af_{\modeindex}^{2} + \frac{1}{c^{2}} (\af_{n}^{2}-\afThresh^{2}) (\atomic\af - \af_{n})^{2}} \nl
	 & = \frac{\decay}{2\pi}\frac{\atomic\crosssection}{\atomic\af}  \sum_{\modeindex} \frac{\modeProfile^{2}}{\crosssection_{\modeindex}}
	\left( \int_{\afThresh}^{\infty}\dd{\af_{n}} B\fn{$\mode\af; \afThresh$} \right),
\end{align}
where
\begin{equation}
	\label{eq:broadening--integrand}
	B\fn{$\modeTransverse\af; \afThresh$}   =
	\af_{n} \frac{c K }{(cK)^{2} + (1-\frac{\afThresh^{2}}{\af_{n}^{2}}) (\atomic\af - \af_{n})^{2}}.
\end{equation}
Now that the induced line broadening is expressed as fraction of polynomials, the integrand can easily be seen to have two key regimes: when the modes are close to atomic resonance and when they're close to a threshold. Of particular interest, is when these two peaks converge (\autoref{fig:broadening--integrand}) and provide a significant boost to individual modes that is not seen in free space. As long as the threshold frequency is far from the atomic frequency however, $|\afThresh-\atomic\af|\gg\gamma$, the integral is dominated by the atomic resonance. For most modes, with low index, the threshold is indeed far below the atomic frequency. In this case,  $(1-\afThresh^{2}/\af_{n}^{2})\approx 1$ and only near-resonant modes, $ \af_{n} \sim \atomic\af$, contribute: \(B\fn{$\modeTransverse\af; \afThresh$} \approx B\fn{$\modeTransverse\af; 0$}\) approximates a Lorentzian with a very narrow width, $cK$.

Such is the case, the spontaneous emission rate is conveniently split into two terms,
\begin{align}
	\label{eq:broadening}
	\statewidth\fn{\positionTransverse} & = \statewidthThZero\fn{\positionTransverse} + \statewidthTh\fn{\positionTransverse} \nl
	                                    & \equiv {\decay} \frac{\atomic\crosssection}{\atomic\af}  \sum_{\modeindex} \frac{\modeProfile^{2}}{2\crosssection_{\modeindex}}
	\left( \frac{1}{\pi} \int_{\afThresh}^{\infty}\dd{\af_{n}} B\fn{$\mode\af; 0$} +\frac{1}{\pi}\int_{\afThresh}^{\infty}\dd{\af_{n}} \left[B\fn{$\mode\af; \afThresh$}- B\fn{$\mode\af; 0$}\right] \right)\;,
\end{align}
where the second term is, in general, negligible except for \textit{threshold effects}, \textit{i.e.} when $\afThresh \approx \atomic\af$ for individual mode branches.

Considering the first term, this, in turn, can also be well approximated by a unit pulse around resonance, allowing the integral to be performed explicitly:
\begin{align}
	\label{eq:bulkGammainHPC}
	\statewidthThZero\fn{\positionTransverse} \simeq {\decay} \, {\atomic\crosssection} \,  \sum_{\modeindex} \frac{\modeProfile^{2}}{2 \crosssection_{\modeindex}} \approx \statewidth \;,
\end{align}
where the spatial dependence of the spontaneous emission rate on the position, $\positionTransverse$,  can be estimated to average out on summing up over many modes. For example, a typical `holey' fiber, $D= \SI{50}{\micro\meter}$, subject to infra-red laser drive, $\wl=\SI{0.78}{\micro\meter}$, can have $\sim \lfloor {2 D}/{\wl} \rfloor \approx 128$ modes along a central axis and so $\sim 10,000$ transverse  modes in total. With so many modes contributing to spontaneous emission, the radiative environment is similar to that of free space when the atoms are not in the near vicinity of the boundary. No significant spatial dependence remains. Since the density of modes tends to that of free space in the large hollow core diameter limit, $\statewidth$ is expected to be on the same order of magnitude as the free space rate, $\decay$. However, while the distribution of mode polarizations is isotropic in free space, the propagation axis of the modes in a fiber breaks this symmetry. Therefore, the spontaneous emission rate depends on the direction of the atomic dipole moment and can be different from $\decay$. Thus, the bulk contribution to the line broadening, $\statewidthThZero$, will be treated as a phenomenological parameter in the model. The contribution of the second term, identified as a threshold effect, we address in the following section.

\subsection{Vacuum threshold effects}

Although the summation over many transverse modes above leads back to familiar, free-space magnitudes, it is noteworthy that single vacuum mode branches significantly modify atomic properties on their own, when  $\atomic\af$ happens to be close to, \textit{i.e.} within $\gamma$, a threshold frequency, $\afThresh$. This is a significant departure from free-space environments and is a feature regardless of fiber diameter. In fact, this  concerns both the broadening and the shift of the atomic resonance due to vacuum modes.

Continuing with the broadening, the conceptual separation of the emission rate, \autoref{eq:broadening}, is further justified numerically: as the individual terms, both `bulk' and `threshold' effects, are now integrable. In the first case, we approximated with a unit pulse and now, in the second term, the high-frequency (UV) divergence of the prior integrals are countered by the subtraction of the \(\afThresh=0\) component. This allows us to consider the relevance of single branches,
\begin{equation}
	\label{eq:state-broadening}
	\statewidthMode\fn{\positionTransverse} = \frac{\decay}{2\pi}\frac{\atomic\crosssection}{\atomic\af}  \frac{\modeProfile^{2}}{\crosssection_{\modeindex}}
	\int_{\afThresh}^{\infty}\dd{\af_{n}} \left[B\fn{$\mode\af; \afThresh$}- B\fn{$\mode\af; 0$}\right]\;,
\end{equation}
with respect to the overall vacuum emission.
\begin{figure}
	\centering
	\includegraphics[width=1\linewidth]{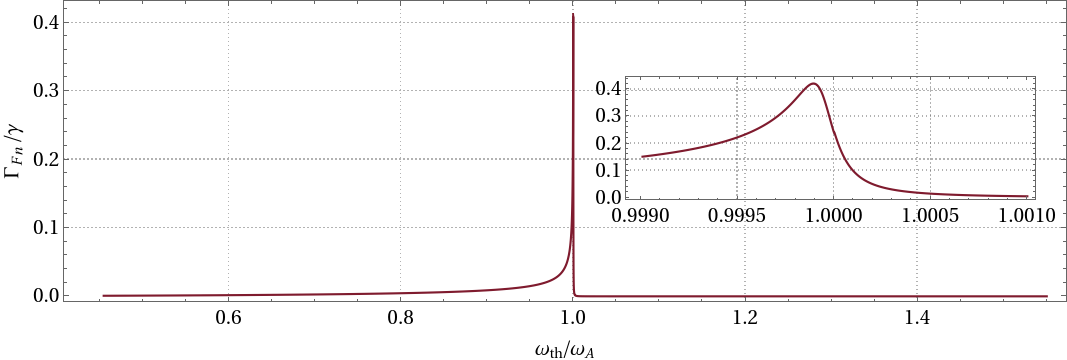}
	\caption{The spontaneous emission of a single, untypical mode. Here we see that a single mode can make a significant contribution to the whole, when the threshold frequency is close to an atomic resonance. This mode was chosen by taking a rough `maximum' value of the allowed mode profiles (0.047), given the beam waist and wavelength. The mean profile value is more stable (actually the same for all modes),i.e. approximately 0.000398.}
	\label{fig:broadening--branch}
\end{figure}
\begin{figure}
	\centering
	\includegraphics[width=1\linewidth]{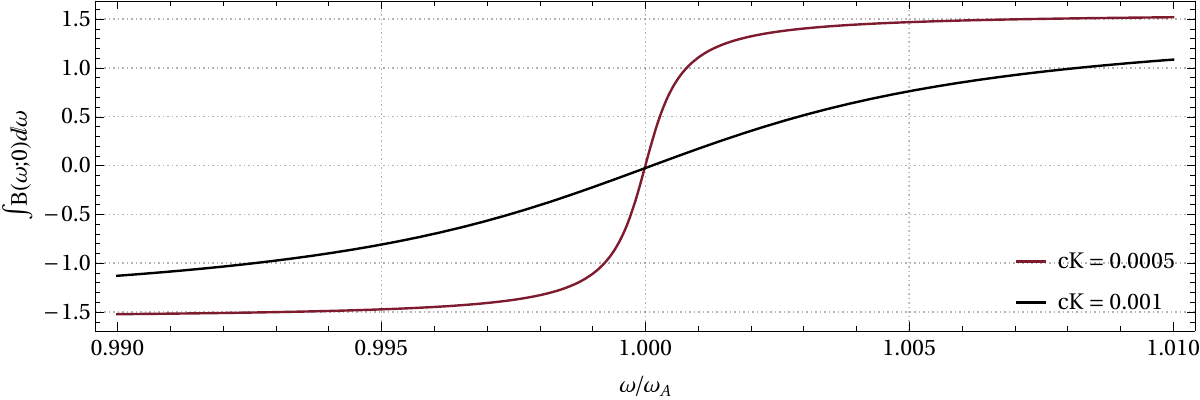}
	\caption{When the fiber threshold is negligible, the broadening function is simply a modified Lorentzian, with location and scaling parameters, that is analytically tractable.}
	\label{fig:broadening--thresholdless}
\end{figure}
%
%In this case, in fact, the removed divergence can be analytically calculated:
%\begin{equation}
%    \int B\fn{$\modeTransverse\af; 0$} d\modeTransverse\af =% 
%        c K \Big(
%        \frac{\atomic\af \arctan\fn{$\frac{\af - \atomic\af}{cK}$}}{cK} + 
%   \half \ln\fn{$(cK)^2 + (\af - \atomic\af)^2$}
%   \Big),
%\end{equation}
%as shown in \autoref{fig:broadening--thresholdless}.

The line shift (\autoref{eq:lineshift}) can be treated in the same way. There is a `bulk' effect associated with the many modes with threshold far below the atomic resonance. This contribution is close to homogeneous and is usually incorporated into the observable atomic frequency, $\atomic\af$, which is thus the renormalized bare frequency. Keeping only the  threshold effect,
\begin{align}
	\label{eq:state-lineshift}
	\lightshift\fn{\positionTransverse} & = \frac{\decay}{2\pi}\frac{\atomic\crosssection}{\atomic\af}   \sum_{\modeindex} \frac{\modeProfile^{2}}{\crosssection_{\modeindex}}
	\int_{\afThresh}^{\infty}\dd{\af_{\modeindex}} \Big[ D\fn{$\mode\af; \afThresh$}- D\fn{$\mode\af; 0$} \Big]
	\text{ and }\nl
	D\fn{$\mode\af; \afThresh$}         & = \frac{\af_{\modeindex}^{2} (\af_{\modeindex}^{2} - \afThresh^{2})^{1/2} (\atomic\af - \af_{\modeindex})}{c^{2} K^{2} \af_{\modeindex}^{2} + (\af_{\modeindex}^{2}-\afThresh^{2}) (\atomic\af - \af_{\modeindex})^{2}}\;.
\end{align}
%
%\subsection{Vacuum threshold effects}
%
\begin{figure}[bt]
	\centering
	\includegraphics[width=\linewidth]{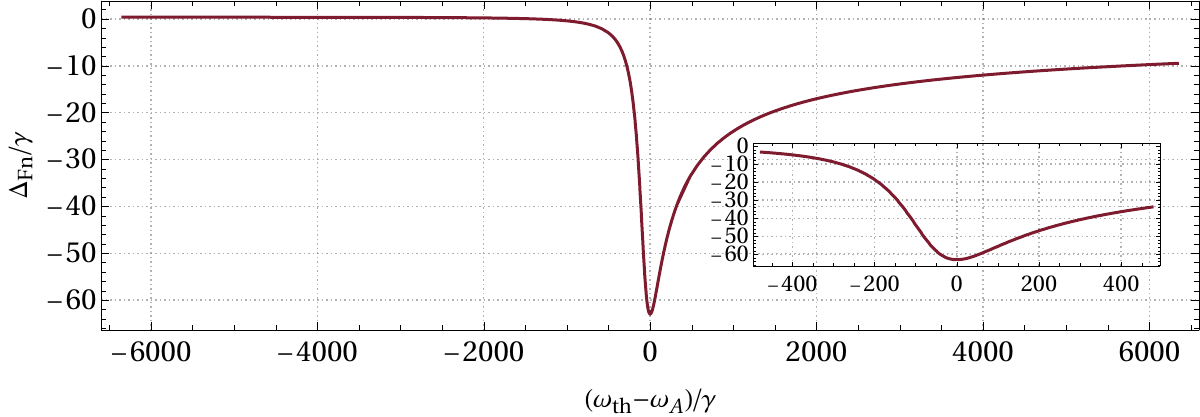}
	\caption{An example of the vacuum-induced line shift (\autoref{eq:state-lineshift}) from a single mode as the threshold frequency is scanned around the vicinity of an atomic resonance: where we have assumed that the fiber waist is \SI{50}{\micro\meter} and that the mode profile value per cross section is 0.05.}
	\label{fig:detuning--threshold}
\end{figure}
Considering then the fiber effects as a whole, it is worth noting that although vacuum-induced line changes are well studied phenomena, in practice they are usually negligible. Particularly, as experimental references often include these effects within the reported numbers. In contrast, we see that thresholding from the fiber geometry, as suggested in \autoref{fig:broadening--branch} and \autoref{fig:detuning--threshold}, can greatly amplify the effect, such that the atomic transition can be both broadened and shifted with respect to free space measurements. Thus, we extend, on a rigorous QED basis, results first suggested in 2007 \cite{szirmai_geometric_2007}, and yet which remain experimentally untested. Furthermore, we note that even the base prefactor above (\autoref{fig:detuning--threshold}) scales inversely with the square of the fiber waist and so even bigger contributions can be expected. This follows from \autoref{eq:state-broadening}:
\begin{equation}
	\frac{\atomic\crosssection}{2\pi}\frac{\modeProfile^{2}}{\crosssection_{\modeindex}} = \frac{3\atomic\wavelength^{2}}{4\pi^{2}\waist^{2}} \cdot \underbrace{\waist^{2}\mode\profile^{2}}_{\text{constant mean}}.
\end{equation}

\subsection{Drive}
\label{sec:drive}
Having discussed a bare fiber system, let us consider the effects of adding the final component: a driving laser. Whereas many modes could be coherently populated,
\[
	\Ham_{\text{dipole}} \simeq \i\hbar \sum_{n} \intInf \modeTime\cre \pauliM \left(\modeProfile \mode\coupling   e^{-\i\Phase} \right)\dd{\wn} + \hc,
\]
we consider a single mode, driven with mean field amplitude, $\alpha$, and displacing quantum fluctuations. There will then be an additional term in the Hamiltonian:
\begin{equation}
	\label{eq:hamiltonian-drive}
	\drive{\Ham} = \pauliM~
	\i\hbar \conjugate{\rabiDrive}
	~\e^{-\i \drivePhase} + \hc,
\end{equation}
where the spatially-dependent Rabi-frequency and phase can be specified for travelling, \( \drive{\rabiTransverse}= \amplitude \drive\coupling \drive\profile\fn{\positionTransverse} \text{ and  } \drivePhaseP=\wn z \), or standing waves, \(\rabiDrive = \amplitude \drive\coupling \drive\profile\fn{\positionTransverse} \cos(\wn z)\text{ and }\drivePhaseP=0\). Here, the Rabi frequency incorporates the dipole coupling constant, the mode function of the driven mode and the drive amplitude itself. The drive phase is considered to be separable: \(\drivePhase = -\drive\af t + \drivePhaseP\). Subsequently, there will be an extra term in the atomic equation of motion (\autoref{eq:evolution--vacuum}:
\begin{align}
	\dt\pauliMRot & = -\pauliMRot \left[ \statewidth\fn{\positionTransverse} + \i \lightshift\fn{\positionTransverse}  \right] -\rabiDrive\e^{\i (\afDiffAtomicDrive\t + \drivePhaseP)}
	-\noiseRot\fn{\t, \v\position},
\end{align}
where in the atomic frame, drive detuning contributes a time dependent phase. Considering the driving frame instead, \( \pauliMRot = \pauliMRotDrive \e^{\i\t \afDiffAtomicDrive} \),
\begin{align}
	\label{eq:evolution--drive}
	\dt\pauliMRotDrive & = -\pauliMRotDrive \left[\statewidth\fn{\positionTransverse} + \i (\afDiffAtomicDrive + \lightshift\fn{\positionTransverse})  \right] -\rabiDrive\e^{\i\drivePhaseP}
	-\noiseRotDrive\fn{\t, \v\position}.
\end{align}

\subsection{Atomic polarization}
Having built up the main equation of motion, we now consider the atomic evolution. Of prime importance, is the steady-state behaviour, but, as we want to consider the noise properties, we first consider the full solution:
\begin{align}
	\label{eq:evolution--general}
	\dt\pauliMRotDrive\fn{t}       & = f \pauliMRotDrive\fn{t} + g\fn{t}\nl
	\implies \pauliMRotDrive\fn{t} & = C \e^{f(t_2 - t_1)} + \e^{f(t_2 - t_1)}\int_{t_1}^{t_2} g\fn{t} \e^{-f(t_2 - t_1)}dt,
\end{align}
where $C$, $f$ and $g$ are arbitrary placeholders and we note that $f$ is not time-dependent. Assuming the first term to be transient, then the atomic polarization follows:
\begin{equation}
	\label{eq:SigmaSteadyState}
	\pauliMRotDrive \simeq \frac{- \rabiDrive\e^{\i\drivePhaseP}}{\statewidth\fn{\positionTransverse} + \i (\afDiffAtomicDrive + \lightshift\fn{\positionTransverse})} + \rotDrive{\Sigma}\fn{\t, \v\position},
\end{equation}
where the first term is the steady-state coherent atomic polarization due to pumping and the second, is the time-integrated white noise.

The mean ground, steady-state, polarization is then
\begin{equation}
	\label{eq:steady-state-number}
	\ev{\pauliPRot\pauliMRot} = \frac{\abs{\rabiTransverse}^{2}}{\statewidth^{2} + (\afDiffAtomicDrive + \lightshift)^{2}}
\end{equation}
and the fluctuations around the steady state are described by the second-order correlation:
\begin{equation}
	\label{eq:SigmaNoise}
	\langle \Sigma(t_1) \Sigma^\dagger(t_2) \rangle \approx \frac{2 \statewidth\fn{\positionTransverse}}{\statewidth\fn{\positionTransverse}^2 + (\afDiffAtomicDrive + \lightshift\fn{\positionTransverse})^2} \, \delta(t_1-t_2).
\end{equation}
Here, the approximation comes from replacing a narrow, peaked, function with the unit pulse on time scales of the order of $\Gamma_F^{-1}$, which is a good approximation for the timescales of atomic motion that we will consider next. It is also worth noting that this correlation function exactly expresses the fluctuations of the atomic polarization, i.e.
\begin{equation}
	\left\langle \left(\pauliMRotDrive (t_1) - \langle \pauliMRotDrive \rangle\right) \left(\pauliMRotDrive^\dagger (t_2) - \langle \pauliMRotDrive^\dagger \rangle\right)  \right\rangle = \langle \Sigma(t_1) \Sigma^\dagger(t_2) \rangle\,.
\end{equation}
With these expressions for the atomic polarization, we can now consider the forces acting on the atomic centre-of-mass motion.
\section{Forces}
\begin{figure}[bt]
	\centering
	\includegraphics[width=0.75\linewidth]{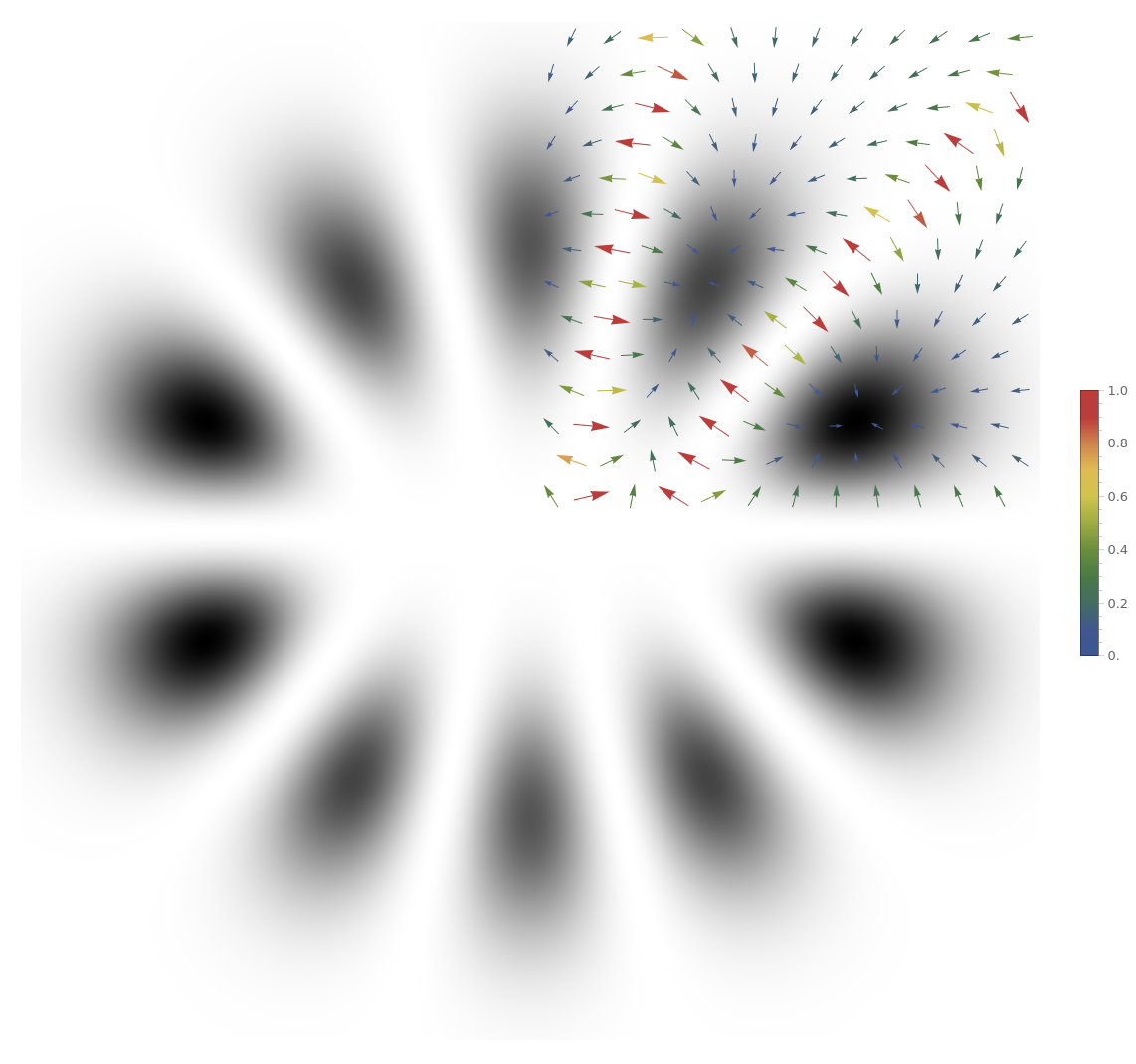}
	\caption{An example of the two-dimensional mode profile gradient, normalised with respect to the longitudinal intensity ($2\grad\drivePhaseP\abs{\rabiDrive}^2$).}
	\label{fig:spatial-gradient}
\end{figure}

The force that an electromagnetic field exerts on a moving atom can be calculated from the commutator of the momentum operator and the total Hamiltonian (\autoref{sec:drive}): such that all modes, driven and in vacuo, contribute. Formally, the commutation amounts to the gradient of the Hamiltonian with respect to the atomic position, \textit{i.e.},
\begin{align}
	\label{eq:force--combined}
	\v{\driveForce} = & - \grad \rotDrive H \nl
	=                 & -
	\underbrace{\i\hbar \sum_{n} \intInf \modeTime\creRotDrive \pauliMRotDrive  \mode\coupling \grad\left(\modeProfile   e^{-\i\Phase } \right)\dd{\wn}}_{vacuum}
	\underbrace{-\pauliMRotDrive \i\hbar \grad\Big( \conjugate\rabiTransverse \e^{-\i \drivePhaseP} \Big)}_{\text{drive}}  + \hc\,
\end{align}
An efficient way to calculate the gradient of the field profiles (\autoref{fig:spatial-gradient}) is thus central to any practical force calculation. And so we turn to our analytical approximation of the mode functions (\autoref{eq:gradient-field-profile}).
With easy access to the spatial derivatives, we can then adapt the standard semiclassical theory of laser cooling \cite{gordon_1980,dalibard_1985a}, applicable to atoms above ultracold temperatures, where the atomic centre--of--mass motion can be considered a classical degree of freedom. The mechanical effects of the field on the atom can then be simulated by means of a classical Langevin equation, which includes a friction force and also the recoil-induced heating effect.

From the general expression above, we will derive three key terms of the Langevin equation. First of all, the quantum mechanical mean, corresponding to the force acting on the atom at rest. Secondly, for an atom moving slowly, the non-adiabatic response of the mean force to linear order in the velocity. This can yield a friction force on the atom. Finally, we will consider heating from quantum noise. The internal atomic polarization, $\pauliMRotDrive$, includes noise sources that give rise to forces with zero mean, but non-vanishing second-order correlation function. By modelling this correlation by means of a classical random force, with a matched diffusion coefficient, we gain a semiclassical view of the heating process.

\subsection{Mean steady-state force}
In a driven system in free space, it is naturally the second term of \autoref{eq:force--combined} that dominates and so the first component that we consider here. The first term of the force, the vacuum component, would only contribute to the recoil heating as photons would be scattered from incoming laser light. In a fiber however, remarkably, the vacuum does not just scatter photons into the other modes, but creates an additional term that survives in the mean. Thus, we must explore both terms, below.

\subsubsection{Drive}
First considering the drive term, we expand \autoref{eq:force--combined} in the steady-state limit,
\begin{align}
	\drive{\ev{\rotDrive{\v\force}}} & = \i\hbar \Big(\grad \conjugate\rabiTransverse  -\i\grad\drivePhaseP \conjugate\rabiTransverse  \Big)
	\frac{\rabiDrive}{ \statewidth\fn{\positionTransverse} + \i (\afDiffAtomicDrive + \lightshift\fn{\positionTransverse})} + \hc \nl
	                                 & = \frac{\hbar}{\statewidth\fn{\positionTransverse}^{2} + (\afDiffAtomicDrive + \lightshift\fn{\positionTransverse})^{2}} \nl
	                                 & \Big[ \statewidth\fn{\positionTransverse} (\grad\drivePhaseP\intensity{\rabiTransverse} + \grad\conjugate\drivePhaseP \intensity{\rabiTransverse}) + (\afDiffAtomicDrive + \lightshift\fn{\positionTransverse}) \grad\intensity{\rabiTransverse} \nl
	                                 & + \i (\statewidth\fn{\positionTransverse}
		(\grad\intensity{\rabiTransverse} - 2\grad\rabiTransverse\conjugate\rabiTransverse)
		+ (\afDiffAtomicDrive + \lightshift\fn{\positionTransverse})
		(\grad\conjugate\drivePhaseP\intensity{\rabiTransverse} - \grad\drivePhaseP\intensity{\rabiTransverse})
		)
		\Big].
\end{align}
And so, in the limit of real Rabi frequencies and phases, the drive force simplifies to
\begin{align}
	\drive{\ev{\rotDrive{\v\force}}} = \frac{ \hbar}{\intensity{\stateLineP}}
	\Big[%&
		%\intensity{\rabiDrive}\grad\lightshift\fn{\positionTransverse} +\nl
		2\grad\drivePhaseP \intensity{\rabiDrive} \Re(\stateLineP) + \grad(\intensity{\rabiDrive})\Im(\stateLineP)
		\Big],
\end{align}
where, for conciseness, we have combined the complex atomic line properties:
\begin{equation}
	\stateLineP =  \statewidth\fn{\positionTransverse}  + \i(\afDiffAtomicDrive + \lightshift\fn{\positionTransverse} ).
\end{equation}
Such an expression will be very familiar to those working throughout atomic physics. The overall form is very similar to that for free-space, with the first term representing the radiation pressure and the second, the gradient force. The fiber modifies these forces only via the general spatial dependence of the mode profiles and through the fiber-induced light-shift and broadening.
\subsubsection{Vacuum field}
It is of little surprise that the mean force of a driving laser, via the mode function, directly reflects the fiber-imposed geometry of the electromagnetic field. What is remarkable, is that there is a further contribution from the vacuum. Here, the mean does not vanish due to a non-trivial mechanical manifestation of the vacuum threshold effect: something specific to a constrained geometry. This can be seen below.

Remaining in the atomic frame of reference, the mean vacuum force is given by
\begin{equation}
	\ev{\rot{\v{\force}}_{\text{vac.}}} = - \i\hbar \sum_{n} \intInf \ev{\mode\creRot \pauliMRot} \grad\left(\modeProfile \mode\coupling   e^{-\i\Phase } \right)\dd{\wn} + \hc
\end{equation}
Expanding the field operator, \autoref{eq:field-evolution}, then reveals two distinct contributions: that from the (atomically) `free' field and that from the atomic `self-reaction'. This latter category then includes the force at branch thresholds where the vacuum-induced AC-Stark shift is considerable:
\begin{align}
	\label{eq:force--vacuum}
	\ev{\rot{\v\force}_{\text{vac.}}} & = \ev{\rot{\v\force}_{\text{free}}} + \ev{\rot{\v\force}_{\text{react}}} \nl
	                                  & = -\i\hbar~  \sum_{\modeindex}\intInf \underbrace{\ev{ \modeTransverse\creRot\fn{\wn,\tzero,\v\position} \pauliMRot}}_{\text{operator}} ~\mode\coupling \grad\Big( \mode\profile \e^{-\i \Phase} \Big)  \underbrace{\e^{\i \afDiff t -\mode\lossFiber t}}_{\text{temporal}}  \dd{\wn} + \hc \\
	                                  & - \i\hbar~ \sum_{\modeindex}\intInf \underbrace{\ev{\pauliPRot\pauliMRot}}_{\text{operator}} \mode\coupling^{2}
	\Big(\grad
	\intensity{\modeTransverse{\profile}}
	- \i\grad\Phase
	\modeTransverse{\profile}^{2}
	\Big) \underbrace{\frac{\mode\lossFiber + \i\afDiff}{\mode\lossFiber^{2} + \afDiff^{2}}}_{\text{not temporal}} \dd{\wn}  + \hc
\end{align}
These terms have quite distinct features. The first term necessarily vanishes when the operator acts on the ground state. The second term, having already temporally integrated over the atom-radiated field with a forgetful (Markovian-like) approximation, can be transformed into an explicitly real form:
\begin{align}
	\label{eq:force-react}
	\ev{\rot{\v\force}_{\text{react}}} & = \hbar~ \sum_{\modeindex}\intInf \ev{\pauliPRot\pauliMRot}~ \frac{\mode\coupling^{2}}{\mode\lossFiber^{2} + \afDiff^{2}} \Big[ \afDiff\grad \intensity{\modeTransverse\profile} - \lossFiber \grad\Phase  \modeTransverse\profile^{2}  \Big] \dd{\wn}.
\end{align}
If we assume only z-dependence in the spatial phase, then the second term of this expression disappears: due to the symmetry of integrating over wavenumbers with both positive and negative components. Furthermore, if we recognise the similarity with \autoref{eq:state-lineshift}, then we arrive at a curiously concise form for the self-reaction:
\begin{equation}
	\label{eq:force--self-reaction}
	\ev{\rot{\v\force}_{\text{react}}} = \hbar~\ev{\pauliPRot\pauliMRot} \grad\lightshift\fn{\positionTransverse}.
\end{equation}
Two things are immediately clear from this expression. First of all, we reiterate that in homogeneous free space there is no such contribution to the mean force, $\lightshift=0$, and so this vacuum effect is intimately related to the geometric confinement, particularly the spectral structure of the modes. Secondly, this unusual contribution is not necessarily negligible; it could in fact be a noticeable contribution to the mean force in threshold domains, i.e. where $\afThresh \approx \atomic\af$. Although such a force only becomes noticeable when the atomic frequency is close to the threshold of a propagating branch, cf. the resonance-like behaviour of the lineshift in \autoref{fig:detuning--threshold}, the fiber induced light shift inherits the spatial variation of the threshold mode, $\lightshift\fn{\positionTransverse} \propto \modeProfile^{2}$, which is typically a strongly modulated, high-index mode of the hollow-core fiber. Thus, we predict an appreciable mean steady state \textit{vacuum reaction force} acting on a laser-driven atom within a hollow-core optical fiber.

\paragraph{}
This new force has no direct equivalent in free, 3-D, space, but we can interpret it as a necessary consequence of the geometry-induced light shift, \autoref{eq:state-lineshift}, where the induced detuning is naturally accompanied by an induced gradient force. It also worth noting however, that this threshold force depends on atomic excitation, and so in practice, on external laser drive. In accordance with \autoref{eq:steady-state-number},  the explicit form of the mean vacuum reaction force, for a driven atom, is thus
\begin{equation}
	\label{eq:foce-react--avg}
	\ev{\rotDrive{\v\force}_{\text{react}}} =
	\hbar~ \frac{\abs{\rabiDrive}^{2} }{\statewidth\fn{\positionTransverse}^{2} + (\afDiffAtomicDrive + \lightshift\fn{\positionTransverse})^{2}} \grad\lightshift\fn{\positionTransverse},
\end{equation}
with the extra quantum noise vanishing as expected: $\ev{\noiseRot^{\dagger}\noiseRot}=0$.

Having discussed the mean, steady-state, forces which are naturally important for mechanical trajectories, we now turn to explicitly thermodynamic effects, \textit{i.e.} to atomic cooling and heating.

\subsection{Cooling}
\subsubsection{Drive}
If atoms are moving quickly enough that their trajectory has to be taken into account, when considering spatially dependent effects, but slowly enough that we can consider a power series expansion in the velocity, then we can turn to the \textit{hydrodynamic} approximation. Therefore, rather than simply setting the total time derivative to zero, as in the steady-state, we consider the transformation
\begin{equation}
	\dt \rightarrow \pdt + \vAtom \cdot\grad,
\end{equation}
while also expanding the atomic coupling in powers of velocity, \emph{i.e.}
\begin{align}
	\dt\ev{\pauliM} \rightarrow (\pdt + \vAtom  \cdot\grad) (\ev{\pauliM}_{v^{0}} + \ev{\pauliM}_{v^{1}} \cdot\vAtom \ldots).
\end{align}
Therefore,  (from \autoref{eq:evolution--drive}),
\begin{equation}
	(\pdt + \vAtom  \cdot\grad) (\ev{\pauliMRotDrive}_{v^{0}} + \ev{\pauliMRotDrive}_{v^{1}} \cdot\vAtom \ldots)  = -(\ev{\pauliMRotDrive}_{v^{0}} + \ev{\pauliMRotDrive}_{v^{1}} \cdot\vAtom \ldots)
	\stateLineP -\rabiDrive\e^{\i\drivePhaseP}.
\end{equation}

The mean steady-state limit is then recovered in the zeroth-order solution:
\begin{equation}
	\ev{\pauliMRotDrive}_{v^{0}} = -\frac{1}{\stateLineP}\drive{\rabiTransverse}\e^{\i\drivePhaseP}.
\end{equation}
For first-order in velocity,
\begin{equation}
	\atomic{\v\velocity} \cdot \grad\ev{\pauliMRotDrive}_{v^{0}} = - \atomic{\v\velocity} \cdot \stateLineP \ev{\pauliMRotDrive}_{v^{1}},
\end{equation}
such that
\begin{equation}
	\ev{\pauliMRotDrive}_{v^{1}} = - \frac{1}{\stateLineP}\grad \ev{\pauliMRotDrive}_{v^{0}}.
\end{equation}
The resulting, velocity-dependent, steady-state solution for a single atom is then
\begin{equation}
	\ev{\pauliMRotDrive} \simeq -\frac{1}{\stateLineP} \rabiDrive \e^{\i\drivePhaseP}
	+ \vAtom \cdot
	\frac{e^{i \drivePhaseP}}{\stateLineP^{2}}  \Big(- \frac{\grad\stateLineP}{\stateLineP} +\i \grad\drivePhaseP  + \grad \Big) \rabiDrive,
\end{equation}
such that the mean velocity-dependent force, in the hydrodynamic limt, follows (\autoref{eq:force--combined}):
\begin{equation}
	\ev{{\v\driveForce}_{v}} = \hbar \Big( \conjugate{\rabiDrive} \grad\drivePhaseP + \i\grad\conjugate{\rabiDrive} \Big)
	\Big( \vAtom \cdot
	\frac{1}{\stateLineP^{2}} \Big[ \Big( \frac{\grad\stateLineP}{\stateLineP} -\i \grad\drivePhaseP  - \grad \Big) \rabiDrive \Big] \Big) + \hc
\end{equation}
When the coefficient of the velocity, $\vAtom$, is negative, this is a friction force and so can be used to cool the atoms. The Doppler-cooling effect, i.e. the one based on the variation of the scattering rate due to the Doppler-shift of the resonance with respect to the drive laser, is included in this term up to linear order in velocity. If we limit ourselves to real-valued field profiles then this simplifies further:
\begin{align}
	\ev{{\v\driveForce}_{v}} = & \hbar \grad\drivePhaseP \Big[ \vAtom \cdot \frac{1}{\stateLineP^{2}} \Big( \frac{\grad\stateLineP}{\stateLineP} - \frac{\grad}{2} - \i \grad\drivePhaseP \Big) \rabiDrive^{2} \Big] \nl
	                           & + \hbar \grad\rabiDrive \Big[  \vAtom \cdot \frac{1}{\stateLineP^{2}} \Big( \grad\drivePhaseP + \i\frac{\grad\stateLineP}{\stateLineP} - \i \grad \Big) \rabiDrive \Big]
	+ \hc,
\end{align}
where we have assumed that the gradient of the atomic velocity and the double gradient of the phase vanish. Unpacking the Hermitian conjugate, the full, real-valued, force can be written as
\begin{align}
	\ev{{\v\driveForce}_{v}} = & -\hbar \grad\drivePhaseP \Big[ \vAtom \cdot \frac{1}{\abs{\stateLine}^{4}} \Big( (\Re(\stateLine)^{2}-\Im(\stateLine)^{2}) \grad\rabiDrive^{2} - 4 \Re(\stateLine)\Im(\stateLine) \rabiDrive^{2} \grad\drivePhaseP  \Big)  \Big]\nl
	                           & + \hbar\grad\rabiDrive \Big[  \vAtom \cdot \frac{1}{\abs{\stateLine}^{4}} \Big(2(\Re(\stateLine)^{2}-\Im(\stateLine)^{2}) \rabiDrive \grad\drivePhaseP - 4 \Re(\stateLine)\Im(\stateLine) \grad\rabiDrive \Big)  \Big]                                                 \\
	                           & + \hbar \grad\drivePhaseP \frac{2\rabiTransverse^{2}}{\abs{\stateLine}^{6}} \vAtom \cdot \Big[ \Big(\Re(\stateLine)+\Im(\stateLine) \Big) \Big( \Re(\stateLine) - \Im(\stateLine) \Big) \grad\abs{\stateLine}^{2}
	- \abs{\stateLine}^{2} \Big( \Re(\stateLine)\grad\Re(\stateLine) - \Im(\stateLine)\grad\Im(\stateLine) \Big) \Big]                                                                                                                                                                                  \\
	                           & +\hbar\grad\rabiTransverse \frac{2\rabiTransverse}{\abs{\stateLine}^{6}} \vAtom \cdot \Big[ 2\Re(\stateLine)\Im(\stateLine) \grad\abs{\stateLine}^2 - \abs{\stateLine}^2 \Big( \Re(\stateLine) \grad\Im(\stateLine) + \Im(\stateLine) \grad\Re(\stateLine) \Big) \Big]
	,
\end{align}
where the last two terms can be neglected away from thresholds. These terms account for the spatial inhomogeneity of the fiber-induced lineshift and broadening. However, the broadening due to the threshold can lead to a cooling mechanism via the Doppler-effect: an increased scattering rate into the threshold branch modes from the drive can yield an enhanced velocity-dependent force which is included up to linear order in velocity in the first term above. This effect has been identified previously as geometrically enhanced cooling in a fibre  in \cite{szirmai_geometric_2007}. Here we presented this effect embedded in a consistent QED theory, accounting fully for the vacuum effects near threshold.

\paragraph{}
For comparison with simpler, more familiar results, we can consider this expression in the limit of a standing wave (no forward momentum) and away from threshold:
\begin{align}
	\ev{{\v\driveForce}} = & \hbar\frac{\Im(\stateLine)}{\abs{\stateLine}^{2}} \grad(\rabiDrive^{2})  \nl
	                       & - 4\hbar\grad\rabiDrive \frac{1}{\abs{\stateLine}^{4}} ~\vAtom \cdot  \Big[  \grad\rabiDrive  \Re(\stateLine)\Im(\stateLine)    \Big].
\end{align}

\subsubsection{Vacuum}
Within a two-level system, the force above carries the biggest contribution to cooling, but, as suggested, the vacuum can also play a role. From \autoref{eq:force--self-reaction}, if we follow a similar procedure in vacuo, assuming real Rabi profiles and phase, then we arrive at
\begin{align}
	\ev{\v{\driveForce}_{v}} \simeq \hbar \lightshift\fn{\positionTransverse} \Big[ & - \vAtom\cdot\grad\rabiDrive^2 \frac{ \Re(\stateLineP)  }{\abs{\stateLineP}^4} \nl
	                                                                                & + 2\vAtom\cdot\grad\drivePhaseP \Im(\stateLineP) \frac{ \rabiDrive^2}{\abs{\stateLineP}^4}\nl
	                                                                                & + 2\vAtom\cdot\grad\Re(\stateLineP)^2 \frac{\rabiDrive^2 }{\abs{\stateLineP}^4} \Big],
\end{align}
to 1st-order in the velocity. Again, we see that in the limit of free space ($\lightshift = 0$, then the vacuum averages to zero. Outside of this limit, close to a threshold frequency, the vacuum contribution is not just non-zero, but potentially an appreciable factor, but which includes both cooling and heating terms, depending on the particular circumstances.

\subsection{Heating}
\label{sec:heating}
As with free-space, even when the component averages vanish, there are still practical effects from the fluctuations. And fibers are no exception. In this context, heating can be treated as a momentum diffusion process induced by a Langevin noise term in the classical equations of motion of the atomic centre-of-mass. The classical Langevin force is defined by means of a diffusion coefficient, $D$, which can be matched to the quantum force auto-correlation function via the relation
\begin{equation}
	\label{eq:diffusion-def}
	\ev{\rot{\v\force}_{\text{free}}\fn{\ensuremath{t_{1}}} \circ \rot{\v\force}_{\text{free}}\fn{\ensuremath{t_{2}}}} \simeq D_{\text{free}}\cdot\delta\fn{\ensuremath{t_{1}-t_{2}}}.
\end{equation}
This correspondence of the recoil diffusion, $D_{\text{free}}$, to the force operator auto-correlation function relies on the possibility of separating different time scales. The force auto-correlation proves to be a narrow peaked function around $t_1-t_2$ on the time scale of the internal dynamics of the electric dipole (i.e., $\Omega^{-1}, \decay^{-1}$). The relevant time scale of the atomic motion is the inverse of the recoil frequency which is much slower than the internal dynamics. Therefore, $t_1$ and $t_2$ are practically identical on this time scale; hence the approximation by a unit pulse, $\delta$, for the description of the fluctuating atomic motion. The coefficient in front of a normalized narrow-peaked function of $t_1-t_2$ defines $D_{\text{free}}$. From the above form, it follows that $D_{\text{free}}$ can be directly obtained as the time integral of the force auto-correlation function.

The force correlation can be taken in the steady-state of \autoref{eq:evolution--drive}, assuming an immobile atom. It can be expanded into an explicit form:
\begin{align}
	\ev{\rot{\v\force}_{\text{free}}\fn{\ensuremath{t_{1}}} \circ \rot{\v\force}_{\text{free}}\fn{\ensuremath{t_{2}}}} & = \hbar^{2} \sum_{\modeindex_{1},\modeindex_{2}}\int_{-\infty}^{+\infty}\dd{\wn_{1}}\int_{-\infty}^{+\infty}\dd{\wn_{2}} ~~\ev{\pauliPRot \commutator{\modeTimes{\annRot}{2}}{\modeTimes{\creRot}{1}} \pauliMRot} \modes{\coupling}{1} \modes{\coupling}{2}~ \nl & ~~~  \grad \Big(\modes{\profile}{2} \e^{\i\Phases{2}} \Big) \circ \grad \Big(\modes{\profile}{1} \e^{-\i\Phases{1}}  \Big)  \e^{-\i(\afDiffs{1}\t_{1}-\afDiffs{2}\t_{2})} \e^{-\lossFiber_{1}\t_{1} -\lossFiber_{2}\t_{2}}.
\end{align}
As the forces are vectors with three spatial components, the correlation is considered simultaneously for all, 3x3, cases by using the dyadic product of vectors (denoted by `$\circ$').  Only the time differences, $t_1-t_2$, are relevant, hence one can freely consider the times around $t_{1,2}\approx 0$ and neglect the exponential decays with $\kappa$. The commutator vanishes but for identical modes, i.e. $\modeindex_{1}=\modeindex_{2}$ and $\wn_{1}=\wn_{2}$, reducing the double sum and integral to single ones. The remaining summation over the modes is broadband and flat around the resonance, $\afDiffs{1} \approx 0$, hence only the region $t_1\approx t_2$ contributes to the correlation. This underlies the form given above in \autoref{eq:diffusion-def}, that the force correlation function is tightly peaked around $t_1= t_2$. The time integral can be performed in the spirit of the Markovian approximation, and one obtains
\begin{multline}
	D_{\text{free}}= 2\pi \hbar^{2} \ev{\pauliPRot\pauliMRot} \sum_{\modeindex}\intInf \mode\coupling^{2}\Big[\Big( \grad \modeProfile \Big) + \Big( \grad\Phase \Big)\modeProfile \Big] \\ \circ \Big[\Big( \grad \modeProfile \Big) + \Big( \grad\Phase \Big)\modeProfile \Big]  \delta\fn{\ensuremath{\afDiff}} \dd{\wn} \nl
	\approx \hbar^{2} {\decay} \ev{\pauliPRot\pauliMRot} \, \sum_{\modeindex} \Big[\Big( \grad \modeProfile \Big) +k_A \hat e_z \modeProfile \Big] \circ \Big[\Big( \grad \modeProfile \Big) +k_A \hat e_z \modeProfile \Big]  \frac{\atomic\crosssection}{\modeTransverse\crosssection}\frac{\atomic\af}{\sqrt{\atomic\af^2-\omega_{th,n}^2}}  .
\end{multline}
where $c k_A=\sqrt{\omega_A^2-\afThresh^2}$. The summation over the transverse modes is similar to what we encountered in the case of the spontaneous emission rate, $\Gamma_F$. That summation basically reduces to mode counting, as all of the transverse branches provide a dissipation channel regardless of their index. In contrast, the gradient here scales differently: the lower the index, the smaller the gradient. Thus, assuming strictly linear dependence of the gradient on the mode index (i.e. that the nodes split the space in the hollow core more or less uniformly), the summation can be carried out:
\begin{equation}
	D_{\text{free}}= \hbar^{2} k_A^2 \,  2 \statewidth \ev{\pauliPRot\pauliMRot} \, \left( \begin{array}{ccc}
			1/3 & 1/4 & 1/2 \\
			1/4 & 1/3 & 1/2 \\
			1/2 & 1/2 & 1
		\end{array}\right),
\end{equation}
where we used $k_A\approx \omega_A/c$.

Another source of heating are the fluctuations of the atomic polarization, described in \autoref{eq:SigmaNoise}, which is translated to a fluctuation of the dipole force given in the second term of \autoref{eq:force--combined}. Simple replacement then leads to a diffusion coefficient:
\begin{align}
	\ev{\op{\v\force}_{\text{dr}}\fn{\ensuremath{t_{1}}} \circ \op{\v\force}_{\text{dr}}\fn{\ensuremath{t_t{2}}}} & \simeq D_{\text{dr}}\cdot\delta\fn{\ensuremath{t_{1}-t_{2}}} \nl
	\therefore D_{\text{dr}}                                                                                      & = \hbar^2 \Big( \frac{1}{|\rabiTransverse|^2}\grad\conjugate\rabiTransverse \circ \grad\rabiTransverse +  \grad\drivePhaseP \circ \grad\drivePhaseP  \Big) 2 \statewidth \ev{\pauliPRot\pauliMRot},
\end{align}
where we used the steady excitation result, \autoref{eq:steady-state-number}.This has a similar form to the recoil heating. While the axial component is equal to that of the recoil diffusion, the gradient in the plane transverse to the fibre axis is typically small. That is, diffusion from the drive force fluctuations in this plane is negligible unless the driven mode has a high mode index, so
\begin{align}
	D_{\text{dr}} & \approx  \hbar^{2} k_A^2 \,  2 \statewidth \ev{\pauliPRot\pauliMRot} \, \left( \begin{array}{ccc}
			                                                                                               0 & 0 & 0 \\
			                                                                                               0 & 0 & 0 \\
			                                                                                               0 & 0 & 1
		                                                                                               \end{array}\right)
\end{align} for small numbers of modes.

Finally, note that the quantum fluctuations of the polarization also lead to heating, via the vacuum force in \autoref{eq:force--self-reaction}. It can be derived from the force by calculating the temporal correlations of $F (t) \propto \pauliPRot\pauliMRot$ while assuming Gaussian noise in $\pauliPRot\pauliMRot$:
\begin{equation}
	D_{\text{react}} \approx \hbar^2
	\frac{2 \statewidth\fn{\positionTransverse}}{\statewidth\fn{\positionTransverse}^2 + (\afDiffAtomicDrive + \lightshift\fn{\positionTransverse})^2} \ev{\pauliPRot\pauliMRot} \grad\lightshift\fn{\positionTransverse}.
\end{equation}
Such is the case, this diffusion coefficient is quadratically small in $|\lightshift|/\gamma$, with respect to the recoil and drive fluctuations. In most practical circumstances therefore, it will be negligible. Conceptually however, in situations optimised for the enhancement of the vacuum threshold effects, this term could be a rich subject of interest.

\section{Practical demonstration}
To illustrate the practicality of the theory above, we consider a simple simulation, as detailed using the Wolfram programming language in \autoref{sec:simulation-details}. Placing a single rubidium atom in the 2D profile of the optical field, we can simulate its behaviour through time, assuming that it is moving slowly enough that Newton's law of motion is appropriate. The optomechanical effects can then be naturally included in the force terms, but where we have to manually implement an RK4 algorithm that allows us to consider changes on the level of a single `step'. This is necessary due to the stochastic nature of the heating.

\begin{figure}[tb]
	\centering
	\subfigure{\includegraphics[width=0.35\textwidth]{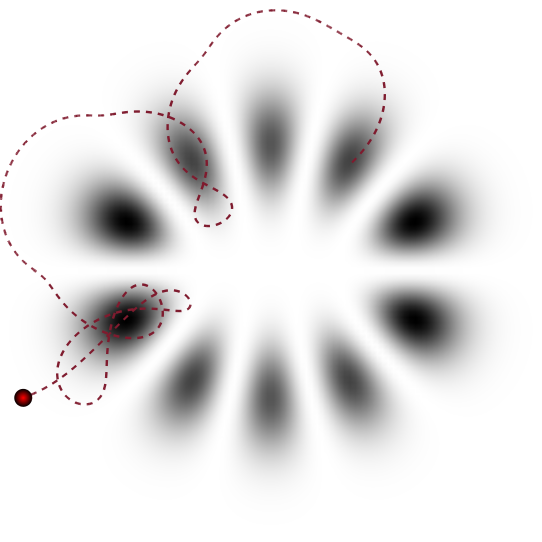}}
	\subfigure{\includegraphics[width=0.643\textwidth]{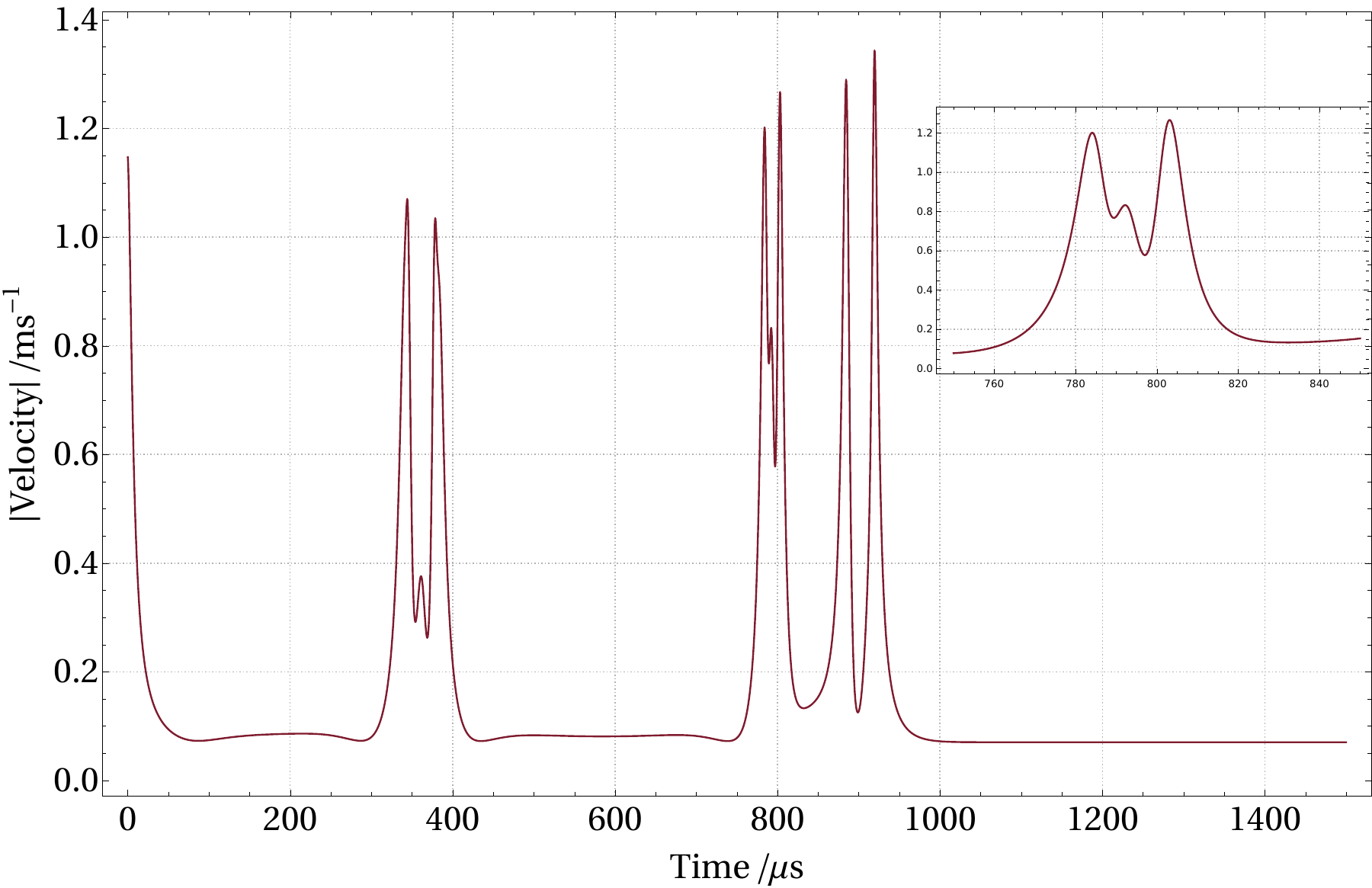}}
	\caption{An example trajectory from a simple semi-classical simulation. Here we consider a particular set of parameters: $\Gamma=37.6991$; $\Gamma_F=37.6991$; $\Delta=7728.32$; $\Delta_F=0.0$; $\lambda_A=0.78$; $\hbar=1$; $m=2.27369$; $w0=10$; $k=8.05537$; $g=2443.94$; $x0=9$; $y0=12$; $vx0=0.812$; $vy0=0.81$; $t=1500$ and $dt=0.05$.}
	\label{fig:simulation-overview}
\end{figure}

\begin{figure}[tb]
	\centering
	\includegraphics[width=\linewidth]{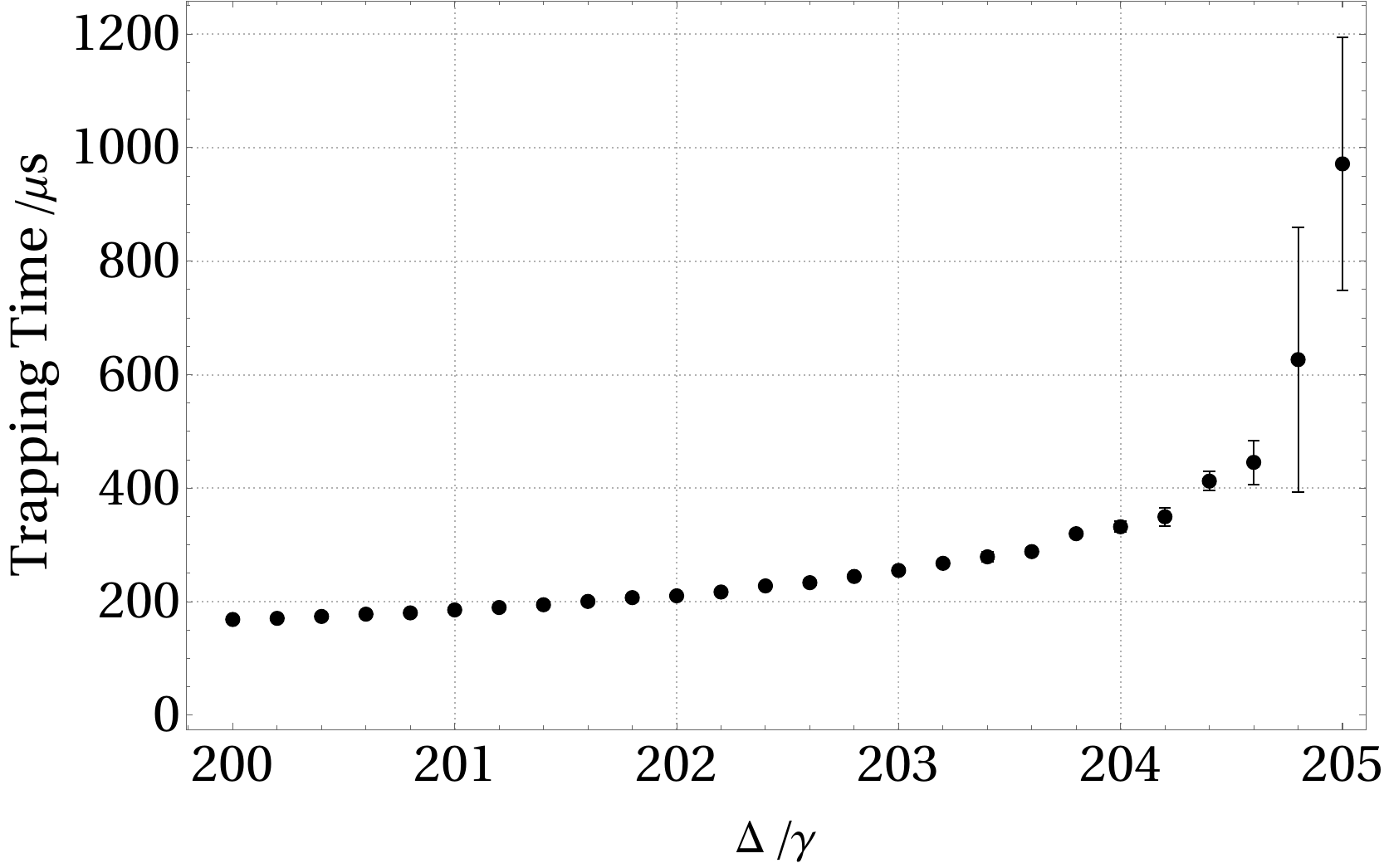}
	\caption{Running the simulation for various detunings and 50 repetitions, we see the stochastic effect of the heating and a credible increase in trapping time with detuning.}
	\label{fig:trapping-time-vs-detuning}
\end{figure}

As can be seen in the figures (\autoref{fig:simulation-overview},\autoref{fig:trapping-time-vs-detuning}), such a simple algorithm can still reveal complicated motions that are a balance of cooling and heating while moving through a relevant, but arbitrary inhomogeneous profile. This is largely efficiently possible because of our analytical approach to the spatial modes, that allows us to compile the spatial distribution and its derivative in advance. In particular, we model limits on the transverse trapping time from spatially varying diffusion, as an example of something that is now easy to simulate while being extremely difficult to analytically predict.

We will explore the particulars of the new vacuum forces however, in another, more targeted paper.

\section{Conclusion}
\label{sec:conclusion}
In conclusion, we have presented a full QED model for a two-level atom travelling slowly through a hollow-core optical fiber. Having introduced a new form for the loss that leads to soluble, consistent equations and an analytical model for efficient calculation of mode profiles and their gradients, justified with respect to numerical simulation, we then constructed a suitable Hamiltonian. Evaluating this Hamiltonian in the context of Heisenberg-Langevin equations, we highlighted the fiber-specific departures from the free-space results: in particular showing appreciable changes to the mechanics that arise purely from the vacuum. Extending the model to include simple laser driving, we investigated the atomic polarization and the resultant forces: both new and familiar. As well as the spectroscopic effects that arise from being close to a geometric threshold, we considered the possible implications of heating and cooling in a driven system and, finally, presented a simple simulation that combined all of these effects and that demonstrated the potential in this approach. We expect these results to provide a foundation for the development of many-atom and higher saturation models that will surely be of interest to the flourishing field of intrafiber light atom interaction: particularly in the context of quantum metrology and the construction of practical sensors. Furthermore, having highlighted the existence of new vacuum-threshold effects and their consequent forces, we hope to consider these effects in more detail: particularly their practical detection and their performance in  multilevel systems.

\section{Supplementary Material}
\label{sec:supplementary}
\subsection{Constants and Scaling}
\label{sec:ConstantsAndScaling}
\begin{figure}[tb]
	\centering
	\includegraphics[width=\linewidth]{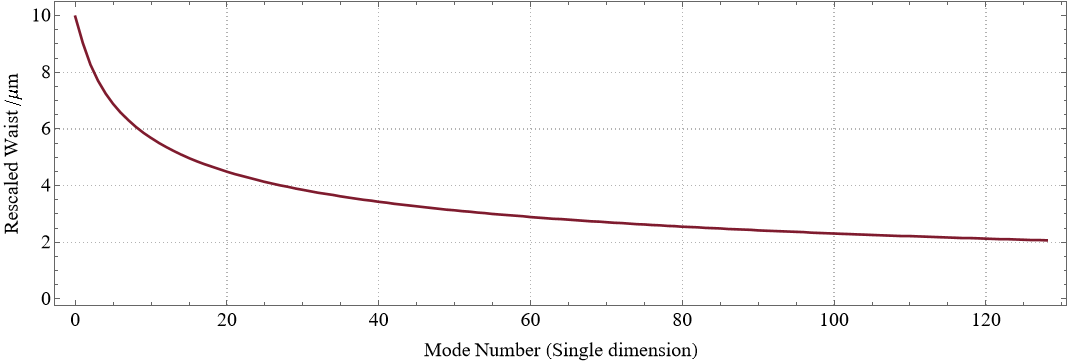}
	\caption{The change in waist needed to maintain a consistent mode cross-section within a $50\mu m$ square as the mode number increases.}
	\label{fig:waist--mode-number}
\end{figure}
For standard, cross-section normalised Hermite-Gaussian beams, the profile maximum has a dependency on the waist which can be neutralised by multiplying by $\waist^2$, \textit{i.e.}
\begin{equation}
	\frac{\atomic\crosssection}{2\pi}\frac{\modeProfile^{2}}{\mode\crosssection}
	= \frac{3\atomic\wl^{2}}{4\pi^{2}\waist^{2}} \cdot \underbrace{\waist^{2}\modeProfile^{2}}_\text{constant mean}.
\end{equation}
For approximating our fiber modes, we do not want the mode cross-section or the profile mean to vary with the mode number and so we artificially alter the waist parameter with increasing mode number. This is done in accordance with \autoref{fig:waist--mode-number}, where the relationship has been investigated numerically.

\subsection{Hermite-Gaussian modes and their derivatives}

The Hermite-Gaussian profiles have the form
\begin{equation}
	\label{eq:hermite-gaussian}
	\text{HG}_{l,m}\fcart =  u_{l}\fn{x,z}u_{m}\fn{y,z} \e^{-\i\wn z},
\end{equation}
where
\begin{equation}
	\label{eq:hermite-gaussian-2d}
	u_{j}\fn{s,z} = \left(\frac{\sqrt{2/\pi}}{2^{j}j!\waist}\right)^{1/2} \left(\frac{q_{0}}{q\fn{z}}\right)^{1/2} \left(-\frac{\conjugate{q}\fn{z}}{q\fn{z}}\right)^{j/2} \hermite{j}{\ensuremath{\frac{\sqrt{2} s}{\width\fn{z}}}} \e^{-\i \frac{\wn s^{2}}{2q\fn{z}}} ~~~~~\forall s \in x,y
\end{equation}
and where we have used the complex beam parameter, $q =z+\i z_R$, and the Rayleigh range, $z_R$.
The transverse derivatives can then be expressed as functions of
\begin{equation}
	\label{eq:hermite-gaussian-derivative}
	\pdv{s}u_{j}\fn{s,z} =  \left(\frac{\sqrt{2/\pi}}{2^{j}j!\waist}\right)^{1/2} \left(\frac{q_{0}}{q\fn{z}}\right)^{1/2} \left(-\frac{\conjugate{q}\fn{z}}{q\fn{z}}\right)^{j/2} \pdv{s}\left[ \hermite{j}{\ensuremath{\frac{\sqrt{2} s}{\width\fn{z}}}} \e^{-\i \frac{\wn s^{2}}{2q\fn{z}}} \right],
\end{equation}
where
\begin{equation}
	\label{eq:hermite-gaussian-derivative-relevant}
	\pdv{s}\left[ \hermite{j}{\ensuremath{\frac{\sqrt{2} s}{\width\fn{z}}}} \e^{-\i \frac{\wn s^{2}}{2q\fn{z}}} \right] =
	\left[\frac{\sqrt{2}}{\width\fn{z}} \frac{2j!}{(j-1)!} \hermite{j-1}{\ensuremath{\frac{\sqrt{2} s}{\width\fn{z}}}}
		- \frac{\i \wn s}{q\fn{z}} \hermite{j}{\ensuremath{\frac{\sqrt{2} s}{\width\fn{z}}}}
		\right] \e^{-\i \frac{\wn s^{2}}{2q\fn{z}}}.
\end{equation}
Therefore,
\begin{equation}
	\label{eq:hermite-gaussian-2D-derivative}
	\pdv{s}u_{j}\fn{s,z} = \frac{2\sqrt{j}}{\width\fn{z}}\left( \frac{-\conjugate{q}\fn{z}}{q\fn{z}} \right)^{1/2}u_{j-1}\fn{s,z} - \i\frac{\wn s}{q\fn{z}} u_{j}\fn{s,z}.
\end{equation}

For $z=0$,
\begin{equation}
	u_{j}\fn{s,0} = \left(\frac{\sqrt{2/\pi}}{2^{j}j!\waist}\right)^{1/2}   \hermite{j}{\ensuremath{\frac{\sqrt{2} s}{\waist}}} \e^{-\frac{s^{2}}{\waist^2}}.
\end{equation}
and
\begin{equation}
	\pdv{s}u_{j}\fn{s,0} = \frac{2}{\waist}\left[ \sqrt{j} u_{j-1}\fn{s,0} - \frac{s}{\waist} u_{j}\fn{s,0} \right]
\end{equation}
Assuming that the fiber version of these modes can be approximated by $\HG{l,m} = u_{l}\fn{x,0} u_{m}\fn{y,0} \e^{-\i\wn z}$, then
\begin{align}
	\grad\HG{l,m} & = \begin{pmatrix}
		                  u_{m}\fn{y,0} \frac{2}{\waist} \left( \sqrt{l} u_{l-1}\fn{x,0} - \frac{x}{\waist} u_{l}\fn{x,0} \right) \\
		                  u_{l}\fn{x,0} \frac{2}{\waist}\left( \sqrt{m} u_{m-1}\fn{y,0} - \frac{y}{\waist} u_{m}\fn{y,0} \right)  \\
		                  u_{l}\fn{x,0} u_{m}\fn{y,0} (-\i\wn)
	                  \end{pmatrix} \e^{-\i\wn z}                    \\
	              & = \begin{pmatrix}
		                  \frac{2}{\waist} \left(\sqrt{l} u_{l-1}\fn{x,0} u_{m}\fn{y,0}     - \frac{x}{\waist} u_{l}\fn{x,0} u_{m}\fn{y,0}   \right) \\
		                  \frac{2}{\waist} \left( \sqrt{m} u_{l}\fn{x,0}   u_{m-1}\fn{y,0} - \frac{y}{\waist} u_{l}\fn{x,0} u_{m}\fn{y,0} \right)    \\
		                  -\i\wn (u_{l}\fn{x,0} u_{m}\fn{y,0})
	                  \end{pmatrix}\e^{-\i\wn z}
	.
\end{align}
Using $H_{n+1}\fn{x} = 2xH_{n}\fn{x} - 2nH_{n-1}\fn{x}$, we can derive a recurrence relation for the single dimension component:
\begin{align}
	u_{j+1}\fn{s,0} & = -(2(j+1))^{-1/2}\Big[ 2j (2j)^{-1/2}u_{j-1}\fn{s,0} - 2(\sqrt{2}\frac{s}{\waist}) u_{j}\fn{s,0} \Big]                          \\
	                & = -(2(j+1))^{-1/2}\sqrt{2}\left[\sqrt{j} u_{j-1}\fn{s,0}  - 2\frac{s}{\waist} u_{j}\fn{s,0} \right]                              \\
	                & = -\frac{1}{\sqrt{1+j}}\left[\sqrt{j} u_{j-1}\fn{s,0}  - 2\frac{s}{\waist} u_{j}\fn{s,0} \right]                                 \\
	                & = -\frac{1}{\sqrt{1+j}}\Big[(\sqrt{j} u_{j-1}\fn{s,0}  - \frac{s}{\waist} u_{j}\fn{s,0}) - \frac{s}{\waist} u_{j}\fn{s,0} \Big].
\end{align}

The gradient of an HG mode can then be expressed in terms of HG modes:
\begin{equation}
	\grad\HG{l,m}\fn{x,y,0} = \begin{pmatrix}
		\frac{2}{\waist}(\sqrt{l} \HG{l-1,m} - \frac{x}{\waist}\HG{l,m}) \\
		\frac{2}{\waist}(\sqrt{m} \HG{l,m-1} - \frac{y}{\waist}\HG{l,m}) \\
		0
	\end{pmatrix}.
\end{equation}
For $z=0$, $\HG{l, m} = \HG{l, m}^*$ and so, from the product rule,
\begin{equation}
	\grad\Big( \intensity{\HG{l,m}} \Big) = 2\HG{l,m} \grad \HG{l,m}.
\end{equation}

\subsection{Simulation details}
\label{sec:simulation-details}
Balancing lab intuition and theoretical relevance, we settled on \si{\micro\meter} and \si{\micro\second} for the basic motion scales and we can define the mass in terms of $\hbar$ from the recoil frequency, $\af_R$, and the wavelength of the D2 transition for Rubidium-87:
\begin{align}
	\af_\text{R,D2} = \frac{\hbar k_{D2}^2}{2 m_\text{Rb}} \nl
	\implies m_{\text{Rb}} = \frac{\hbar \wn_{D2}^2}{2 \af_\text{R,D2}} \nl
	m_{\text{Rb}} \simeq 2.7369.
\end{align}
\begin{lstlisting}[language=mathematica]
noises = {};
forces = {};
forceNs = {};
rs = {};
RK4Step[forceF_, noiseF_, r_, v_, params_] := 
 Module[{k1, k2, k3, k4, F, noiseStep, tStep, M},
  (*r and v are multi-dimensional vectors*)
  tStep = dt /. params;
  M  = m /. params;
  
  noiseStep = noiseF[r, tStep, forceF[r, v]];
 
  F[rr_, vv_] := forceF[rr, vv] + noiseStep;
  k1 = tStep*{v, F[r, v]/M};
  k2 = tStep*{
     v + 0.5*k1[[2]],
     F[r + 0.5*k1[[1]], v + 0.5*k1[[2]]]/M
     };
  k3 = tStep*{
     v + 0.5*k2[[2]],
     F[r + 0.5*k2[[1]], v + 0.5*k2[[2]]]/M
     };
  k4 = tStep*{
     v + k3[[2]],
     F[r + k3[[1]], v + k3[[2]]]/M
     };
  
  {r, v} + (k1 + 2*k2 + 2*k3 + k4)/6.
  ]

simulate[ params_] := Module[{
   steps, positions, velocities,
   xx, yy, vxx, vyy},
  (*---SETUP compiled functions---*)
  ffv[{xx_, yy_}, {vxx_, 
     vyy_}] := (forceMain[{x, y}, {vx, vy}] + 
       forceFriction[{x, y}, {vx, vy}]) /. {x -> xx, y -> yy, 
      vx -> vxx, vy -> vyy} /. params;
  forceCompiled = Compile[
    {{x, _Real}, {y, _Real}, {vx, _Real}, {vy, _Real}},
     Evaluate@ffv[{x, y}, {vx, vy}]];
  ffc[r_, v_] := forceCompiled @@ Flatten[{r, v}];
    
  variance[r_, dt_, 
    f_] := (diffusion[r][[1 ;; 2, 1 ;; 2]]*dt) . (unit@f) /. params;
  varianceCompiled = Compile[
    {{rx, _Real}, {ry, _Real},
     {dt, _Real},
     {fx, _Real}, {fy, _Real}}, 
    Evaluate@variance[{rx, ry}, dt, {fx, fy}]];
  varianceC[r_, dt_, f_] := varianceCompiled @@ Flatten@{r, dt, f};
  
  noise[r_, dt_, f_] := If[Norm@f < 1*^-20,
    {0.0, 0.0},
    RandomVariate[
       UniformDistribution[{-Sqrt[3.] \[Sqrt]Abs@#, 
         Sqrt[3.] \[Sqrt]Abs@#}]] & /@ varianceC[r, dt, f]
    ];
  
  (*---SETUP Simulation proper---*)
  steps = Floor[t/dt] /. params;
  
  positions = ConstantArray[0, steps];
  velocities = ConstantArray[0, steps];
  
  xx = x0 /. params; yy = y0 /. params; vxx = vx0 /. params; 
  vyy = vy0 /. params;

  Do[positions[[i]] = {xx, yy};
   velocities[[i]] = {vxx, vyy};
   
   {{xx, yy}, {vxx, vyy}} = 
    RK4Step[ffc, noise, {xx, yy}, {vxx, vyy}, params];,
   {i, steps}];
  {positions, velocities}
  ]
\end{lstlisting}

\section{Acknowledgements}
	This research was supported by the Ministry of Culture and Innovation and the National Research, Development and Innovation Office within the Quantum Information National Laboratory of Hungary (Grant No. 2022-2.1.1-NL-2022-00004). It was also supported by the H2020-FETOPEN-2018-2020 project, CRYST$^3$, Grant No. 964531.

\bibliography{cryst3}

\begin{thebibliography}{10}
\newcommand{\enquote}[1]{``#1''}

\bibitem{knight_all-silica_1996}
J.~C. Knight, T.~A. Birks, P.~S.~J. Russell, and D.~M. Atkin,
  \enquote{All-silica single-mode optical fiber with photonic crystal
  cladding,} {\protect\JournalTitle{Optics Letters}} \textbf{21}, 1547--1549
  (1996).

\bibitem{cregan_single-mode_1999}
R.~F. Cregan and e.~al, \enquote{Single-mode photonic band gap guidance of
  light in air,} {\protect\JournalTitle{Science}} \textbf{285}, 1537--1539
  (1999).

\bibitem{birks_endlessly_1997}
T.~A. Birks, J.~C. Knight, and P.~S.~J. Russell, \enquote{Endlessly single-mode
  photonic crystal fiber,} {\protect\JournalTitle{Optics Letters}} \textbf{22},
  961--963 (1997). Publisher: Optica Publishing Group.

\bibitem{deutsch_photonic_1995}
I.~H. Deutsch, R.~J.~C. Spreeuw, S.~L. Rolston, and W.~D. Phillips,
  \enquote{Photonic band gaps in optical lattices,}
  {\protect\JournalTitle{Physical Review A}} \textbf{52}, 1394--1410 (1995).

\bibitem{knight_photonic_1998}
J.~C. Knight, J.~Broeng, T.~A. Birks, and P.~S.~J. Russell, \enquote{Photonic
  {Band} {Gap} {Guidance} in {Optical} {Fibers},}
  {\protect\JournalTitle{Science}} \textbf{282}, 1476--1478 (1998). Publisher:
  American Association for the Advancement of Science.

\bibitem{temelkuran_wavelength-scalable_2002}
B.~Temelkuran, S.~D. Hart, G.~Benoit, J.~D. Joannopoulos, and Y.~Fink,
  \enquote{Wavelength-scalable hollow optical fibres with large photonic
  bandgaps for {CO2} laser transmission,} {\protect\JournalTitle{Nature}}
  \textbf{420}, 650--653 (2002). Number: 6916 Publisher: Nature Publishing
  Group.

\bibitem{benabid_compact_2005}
F.~Benabid, F.~Couny, J.~C. Knight, T.~A. Birks, and P.~S.~J. Russell,
  \enquote{Compact, stable and efficient all-fibre gas cells using hollow-core
  photonic crystal fibres,} {\protect\JournalTitle{Nature}} \textbf{434},
  488--491 (2005).

\bibitem{bajcsy_laser-cooled_2011}
M.~Bajcsy, S.~Hofferberth, T.~Peyronel, V.~Balic, Q.~Liang, A.~S. Zibrov,
  V.~Vuletic, and M.~D. Lukin, \enquote{Laser-cooled atoms inside a hollow-core
  photonic-crystal fiber,} {\protect\JournalTitle{Physical Review A}}
  \textbf{83}, 063830 (2011). Publisher: American Physical Society.

\bibitem{christensen_trapping_2008}
C.~A. Christensen, S.~Will, M.~Saba, G.-B. Jo, Y.-I. Shin, W.~Ketterle, and
  D.~Pritchard, \enquote{Trapping of ultracold atoms in a hollow-core photonic
  crystal fiber,} {\protect\JournalTitle{Physical Review A}} \textbf{78},
  033429 (2008).

\bibitem{kepler_cometis_1619}
J.~Kepler, \emph{De cometis libelli tres} (1619).

\bibitem{maxwell_treatise_1873}
J.~C. Maxwell, \emph{A treatise on electricity and magnetism}, vol.~1 (Oxford:
  Clarendon Press, 1873).

\bibitem{ashkin_acceleration_1970}
A.~Ashkin, \enquote{Acceleration and {Trapping} of {Particles} by {Radiation}
  {Pressure},} {\protect\JournalTitle{Physical Review Letters}} \textbf{24},
  156--159 (1970). Publisher: American Physical Society.

\bibitem{phillips_laser_1982}
W.~D. Phillips and H.~Metcalf, \enquote{Laser {Deceleration} of an {Atomic}
  {Beam},} {\protect\JournalTitle{Physical Review Letters}} \textbf{48},
  596--599 (1982). Publisher: American Physical Society.

\bibitem{chu_three-dimensional_1985}
S.~Chu, L.~Hollberg, J.~E. Bjorkholm, A.~Cable, and A.~Ashkin,
  \enquote{Three-dimensional viscous confinement and cooling of atoms by
  resonance radiation pressure,} {\protect\JournalTitle{Physical Review
  Letters}} \textbf{55}, 48--51 (1985). Publisher: American Physical Society.

\bibitem{raab_trapping_1987}
E.~L. Raab, M.~Prentiss, A.~Cable, S.~Chu, and D.~E. Pritchard,
  \enquote{Trapping of {Neutral} {Sodium} {Atoms} with {Radiation} {Pressure},}
  {\protect\JournalTitle{Physical Review Letters}} \textbf{59}, 2631--2634
  (1987). Publisher: American Physical Society.

\bibitem{soding_short-distance_1997}
J.~Söding, R.~Grimm, Y.~B. Ovchinnikov, P.~Bouyer, and C.~Salomon,
  \enquote{Short-{Distance} {Atomic} {Beam} {Deceleration} with a {Stimulated}
  {Light} {Force},} {\protect\JournalTitle{Physical Review Letters}}
  \textbf{78}, 1420--1423 (1997).

\bibitem{alge_all_1997}
W.~Alge, K.~Ellinger, H.~Stecher, K.~M. Gheri, and H.~Ritsch, \enquote{All
  optical local cooling and trapping in {3D},}
  {\protect\JournalTitle{Europhysics Letters (EPL)}} \textbf{39}, 491--496
  (1997).

\bibitem{horak_cavity-induced_1997}
P.~Horak, G.~Hechenblaikner, K.~M. Gheri, H.~Stecher, and H.~Ritsch,
  \enquote{Cavity-{Induced} {Atom} {Cooling} in the {Strong} {Coupling}
  {Regime},} {\protect\JournalTitle{Physical Review Letters}} \textbf{79},
  4974--4977 (1997). Publisher: American Physical Society.

\bibitem{domokos_semiclassical_2001}
P.~Domokos, P.~Horak, and H.~Ritsch, \enquote{Semiclassical theory of
  cavity-assisted atom cooling,} {\protect\JournalTitle{Journal of Physics B:
  Atomic, Molecular and Optical Physics}} \textbf{34}, 187 (2001).

\bibitem{maunz_cavity_2004}
P.~Maunz, T.~Puppe, I.~Schuster, N.~Syassen, P.~W.~H. Pinkse, and G.~Rempe,
  \enquote{Cavity cooling of a single atom,} {\protect\JournalTitle{Nature}}
  \textbf{428}, 50--52 (2004). Publisher: Nature Publishing Group.

\bibitem{murr_large_2006}
K.~Murr, \enquote{Large {Velocity} {Capture} {Range} and {Low} {Temperatures}
  with {Cavities},} {\protect\JournalTitle{Physical Review Letters}}
  \textbf{96}, 253001 (2006).

\bibitem{ghosh_low-light-level_2006}
S.~Ghosh, A.~R. Bhagwat, C.~K. Renshaw, S.~Goh, A.~L. Gaeta, and B.~J. Kirby,
  \enquote{Low-{Light}-{Level} {Optical} {Interactions} with {Rubidium} {Vapor}
  in a {Photonic} {Band}-{Gap} {Fiber},} {\protect\JournalTitle{Physical Review
  Letters}} \textbf{97}, 023603 (2006).

\bibitem{le_kien_light-induced_2006}
F.~Le~Kien, V.~I. Balykin, and K.~Hakuta, \enquote{Light-induced force and
  torque on an atom outside a nanofiber,} {\protect\JournalTitle{Physical
  Review A}} \textbf{74}, 033412 (2006).

\bibitem{yang_atomic_2007}
W.~Yang, D.~B. Conkey, B.~Wu, D.~Yin, A.~R. Hawkins, and H.~Schmidt,
  \enquote{Atomic spectroscopy on a chip,} {\protect\JournalTitle{Nature
  Photonics}} \textbf{1}, 331--335 (2007).

\bibitem{vorrath_efficient_2010}
S.~Vorrath, S.~A. Möller, P.~Windpassinger, K.~Bongs, and K.~Sengstock,
  \enquote{Efficient guiding of cold atoms through a photonic band gap fiber,}
  {\protect\JournalTitle{New Journal of Physics}} \textbf{12}, 123015 (2010).

\bibitem{langbecker_highly_2018}
M.~Langbecker, R.~Wirtz, F.~Knoch, M.~Noaman, T.~Speck, and P.~Windpassinger,
  \enquote{Highly controlled optical transport of cold atoms into a hollow-core
  fiber,} {\protect\JournalTitle{New Journal of Physics}} \textbf{20}, 083038
  (2018).

\bibitem{sommer_laser_2019}
C.~Sommer, N.~Y. Joly, H.~Ritsch, and C.~Genes, \enquote{Laser refrigeration of
  gas filled hollow-core fibres,} {\protect\JournalTitle{AIP Advances}}
  \textbf{9}, 105213 (2019).

\bibitem{wang_enhancing_2022}
Y.~Wang, S.~Chai, T.~Billotte, Z.~Chen, M.~Xin, W.~S. Leong, F.~Amrani,
  B.~Debord, F.~Benabid, and S.-Y. Lan, \enquote{Enhancing fiber atom
  interferometer by in-fiber laser cooling,} {\protect\JournalTitle{Physical
  Review Research}} \textbf{4}, L022058 (2022).

\bibitem{terradas-brianso_ultrastrong_2022}
S.~Terradas-Briansó, C.~A. González-Gutiérrez, F.~Nori, L.~Martín-Moreno,
  and D.~Zueco, \enquote{Ultrastrong waveguide {QED} with giant atoms,}
  (2022). Number: arXiv:2205.07915 arXiv:2205.07915 [quant-ph].

\bibitem{domokos_anomalous_2004}
P.~Domokos, A.~Vukics, and H.~Ritsch, \enquote{Anomalous {Doppler}-{Effect} and
  {Polariton}-{Mediated} {Cooling} of {Two}-{Level} {Atoms},}
  {\protect\JournalTitle{Physical Review Letters}} \textbf{92}, 103601 (2004).

\bibitem{szirmai_geometric_2007}
G.~Szirmai and P.~Domokos, \enquote{Geometric {Resonance} {Cooling} of
  {Polarizable} {Particles} in an {Optical} {Waveguide},}
  {\protect\JournalTitle{Physical Review Letters}} \textbf{99}, 213602 (2007).

\bibitem{cucinotta_perfectly_1999}
A.~Cucinotta, G.~Pelosi, S.~Selleri, L.~Vincetti, and M.~Zoboli,
  \enquote{Perfectly matched anisotropic layers for optical waveguide analysis
  through the finite-element beam-propagation method,}
  {\protect\JournalTitle{Microwave and Optical Technology Letters}}
  \textbf{23}, 67--69 (1999).

\bibitem{selleri_complex_2001}
S.~Selleri, L.~Vincetti, A.~Cucinotta, and M.~Zoboli, \enquote{Complex {FEM}
  modal solver of optical waveguides with {PML} boundary conditions,}
  {\protect\JournalTitle{Optical and Quantum Electronics}} \textbf{33},
  359--371 (2001).

\bibitem{bandres_incegaussian_2004}
M.~A. Bandres and J.~C. Gutiérrez-Vega, \enquote{Ince–{Gaussian} modes of
  the paraxial wave equation and stable resonators,}
  {\protect\JournalTitle{Journal of the Optical Society of America A}}
  \textbf{21}, 873 (2004).

\end{thebibliography}
\end{document}